\documentclass[superscriptaddress,nofootinbib,11pt]{revtex4-1}
\usepackage{amssymb,graphicx,amsmath,verbatim, hyperref}

\newcommand{\beq}{\begin{eqnarray}}
\newcommand{\eeq}{\end{eqnarray}}

\newcommand{\bpm}{\begin{pmatrix}}
\newcommand{\epm}{\end{pmatrix}}
\newcommand{\Z}{\mathbb{Z}}
\newcommand{\R}{\mathbb{R}}
\newcommand{\C}{\mathbb{C}}

\newcommand{\ba}{\left(\begin{array}}
\newcommand{\ea}{\end{array} \right)}
\newcommand{\diag}{{\rm diag}\,}

\usepackage[usenames]{color}

\begin{document}

\title{Lattice $\C P^{N-1}$ model with ${\mathbb Z}_{N}$ twisted boundary condition: \\
bions, adiabatic continuity and pseudo-entropy
}

\author{Toshiaki Fujimori}
\email{toshiaki.fujimori018(at)gmail.com}
\address{Department of Physics, and Research and 
Education Center for Natural Sciences, 
Keio University, 4-1-1 Hiyoshi, Yokohama, Kanagawa 223-8521, Japan}

\author{Etsuko Itou}
\email{itou(at)yukawa.kyoto-u.ac.jp}
\address{Department of Physics, and Research and 
Education Center for Natural Sciences, 
Keio University, 4-1-1 Hiyoshi, Yokohama, Kanagawa 223-8521, Japan}
\address{Department of Mathematics and Physics, Kochi University, Kochi 780-8520}
\address{Research Center for Nuclear Physics (RCNP), Osaka University, Osaka 567-0047}

\author{Tatsuhiro Misumi}
\email{misumi(at)phys.akita-u.ac.jp}
\address{Department of Mathematical Science, Akita University,  Akita 010-8502, Japan}
\address{Department of Physics, and Research and 
Education Center for Natural Sciences, 
Keio University, 4-1-1 Hiyoshi, Yokohama, Kanagawa 223-8521, Japan}

\author{\\Muneto Nitta}
\email{nitta(at)phys-h.keio.ac.jp}
\address{Department of Physics, and Research and 
Education Center for Natural Sciences, 
Keio University, 4-1-1 Hiyoshi, Yokohama, Kanagawa 223-8521, Japan}

\author{Norisuke Sakai}
\email{norisuke.sakai(at)gmail.com}
\address{Department of Physics, and Research and 
Education Center for Natural Sciences, 
Keio University, 4-1-1 Hiyoshi, Yokohama, Kanagawa 223-8521, Japan}

\begin{abstract}
We investigate the lattice $\C P^{N-1}$ sigma model 
on $S_{s}^{1}$(large) $\times$ $S_{\tau}^{1}$(small) 
with the ${\mathbb Z}_{N}$ symmetric twisted boundary condition, 
where a sufficiently large ratio of the circumferences ($L_{s}\gg L_{\tau}$) 
is taken to approximate $\mathbb R \times S^1$. 
We find that the expectation value of the Polyakov loop, 
which is an order parameter of the $\Z_N$ symmetry, 
remains consistent with zero ($|\langle P\rangle|\sim 0$) 
from small to relatively large inverse coupling $\beta$ (from large to small $L_{\tau}$). 
As $\beta$ increases, the distribution of the Polyakov loop on the complex plane, 
which concentrates around the origin for small $\beta$, 
isotropically spreads and forms a regular $N$-sided-polygon shape (e.g. pentagon for $N=5$), 
leading to $|\langle P\rangle| \sim 0$.
By investigating the dependence of the Polyakov loop on $S_{s}^{1}$ direction, 
we also verify the existence of fractional instantons and bions, 
which cause tunneling transition between the classical $N$ vacua and stabilize the ${\mathbb Z}_{N}$ symmetry.
Even for quite high $\beta$, we find that a regular-polygon shape of the Polyakov-loop distribution, 
even if it is broken, tends to be restored 
and $|\langle P\rangle|$ gets smaller as the number of samples increases.
To discuss the adiabatic continuity of the vacuum structure from another viewpoint, 
we calculate the $\beta$ dependence of ``pseudo-entropy" density $\propto\langle T_{xx}-T_{\tau\tau}\rangle$. 
The result is consistent with the absence of a phase transition between large and small $\beta$ regions.
\end{abstract}

\maketitle

\tableofcontents

\newpage


\section{Introduction}

The ${\mathbb C}P^{N-1}$ sigma model \cite{Eichenherr:1978qa,Witten:1978bc,DAdda:1978vbw,DAdda:1978dle}
in two dimensions shares several properties with QCD in four dimensions,
such as asymptotic freedom, instantons, confinement and the generation of a mass gap. 
Such similarities can be explained by several physical setups, 
in which the two-dimensional ${\mathbb C}P^{N-1}$ sigma model effectively describes
various physical properties of four-dimensional gauge theories;
non-Abelian vortices in the non-Abelian gauge-Higgs models 
\cite{Hanany:2003hp,Auzzi:2003fs,Eto:2005yh,Eto:2006cx,Eto:2006pg,Tong:2005un,Eto:2006pg,Shifman:2007ce}
and dense QCD \cite{Nakano:2007dr,Eto:2009bh,Eto:2009tr,Eto:2013hoa}, 
long strings in Yang-Mills theories \cite{Aharony:2013ipa}, 
and an appropriately compactified Yang-Mills theory \cite{Yamazaki:2017ulc}. 
Non-perturbative properties of the ${\mathbb C}P^{N-1}$ model have long been studied analytically 
by the gap equations with the large-$N$ approximation
\cite{Witten:1978bc,DAdda:1978vbw,DAdda:1978dle,Monin:2015xwa,Monin:2016vah,Bolognesi:2019rwq,Flachi:2019jus,Nitta:2017uog,Nitta:2018lnn,Nitta:2018yen,Yoshii:2019yln,Hong:1994uv,Hong:1994te,Milekhin:2012ca,Bolognesi:2016zjp,Milekhin:2016fai,Betti:2017zcm,Flachi:2017xat,Bolognesi:2018njt,Chernodub:2019nct,Pavshinkin:2019bed}
and by lattice simulations \cite{Berg:1981er,Campostrini:1992ar,Farchioni:1993jd,Alles:2000sc,Kataoka:2010sh,Flynn:2015uma,Bruckmann:2015sua,Bruckmann:2016txt,Abe:2018loi,Bruckmann:2018rra,Bonanno:2018xtd,Fujimori:2019skd,Misumi:2019upg,Berni:2019bch}.
In the previous work \cite{Fujimori:2019skd,Misumi:2019upg} of the present authors, 
they have studied the ${\mathbb C}P^{N-1}$ model on 
$S_{s}^{1}$(large) $\times$ $S_{\tau}^{1}$(small) by lattice Monte Carlo simulations. 
A sufficiently large ratio of the circumferences is taken to approximate 
the model on $\mathbb R \times S^1$ with a periodic boundary condition (PBC). 
By adopting the expectation value of the Polyakov loop 
as a confinement-deconfinement order parameter, 
it was shown that its dependence on the compactification circumference undergoes a crossover 
and the peak of its susceptibility gets sharper as $N$ increases. 

The model on $\mathbb{R} \times S^1$ 
with $\Z_{N}$ symmetric twisted boundary conditions (${\mathbb Z}_{N}$-TBC) 
also attracts a lot of attention since the $\Z_{N}$ symmetry, 
whose order parameter is the expectation value of the Polyakov loop operator, 
is exact in this model.
This model admits fractional instantons 
\cite{Eto:2004rz,Eto:2006mz,Eto:2006pg,Bruckmann:2007zh,Brendel:2009mp}, namely instantons with fractionally quantized topological charge,  
typically $1/N$ quantized one for the ${\mathbb Z}_{N}$-TBC 
(see Refs.~\cite{Nitta:2014vpa, Nitta:2015tua, Itou:2018wkm} for 
fractional instantons in other models 
with TBC). 
Then, it has been conjectured 
\cite{Dunne:2012ae,Dunne:2012zk,Sulejmanpasic:2016llc,Tanizaki:2017qhf} 
that the $\Z_{N}$-symmetric vacuum 
of the model on $\mathbb{R} \times S^1$ 
is continuously connected to that on $\mathbb{R}^{2}$
due to the tunneling transition by fractional instantons. 
If there exists such an adiabatic continuity of the vacuum, 
it gives us a deeper understanding of the model on $\mathbb{R}^{2}$ through the weak coupling analysis on $\R \times S^{1}$.
In particular, the nontrivial relation 
between perturbative and nonperturbative contributions,
called ``the resurgent structure"
\cite{Marino:2006hs,Marino:2007te,Marino:2008ya,Marino:2008vx,Argyres:2012vv,Argyres:2012ka,Cherman:2013yfa,Cherman:2014ofa,Behtash:2015kna,Behtash:2015zha,Behtash:2015loa}, 
which has been intensively studied in this model \cite{Dunne:2012ae,Dunne:2012zk,
Misumi:2014jua,
Misumi:2014bsa,
Misumi:2015dua,
Misumi:2016fno,
Buividovich:2015oju,Fujimori:2016ljw,Fujimori:2017oab,Fujimori:2017osz,Dorigoni:2017smz,Fujimori:2018kqp,Ishikawa:2019tnw,Yamazaki:2019arj,Ishikawa:2020eht,Morikawa:2020agf}, 
is expected to play an important role 
in understanding the relation between the weak and strong coupling regime of the model. 

From this perspective, it is of great importance to study 
the ${\mathbb C}P^{N-1}$ model on $\mathbb{R} \times S^1$
with the $\Z_{N}$ symmetric twisted boundary condition, 
with particular attentions to the $\Z_{N}$ symmetry and its order parameter, 
namely the expectation value of the Polyakov loop $\langle P\rangle$. 

In this paper, we investigate the ${\mathbb C}P^{N-1}$ model on $S_{s}^{1}$(large) $\times$ $S_{\tau}^{1}$(small) 
with the $\Z_N$ symmetric twisted boundary condition by lattice Monte Carlo simulations. 
The ratio of the circumferences $L_{s}/ L_{\tau}$ is taken to be sufficiently large
so that the model on $S_{s}^{1} \times S_{\tau}^{1}$ approximately describes that on $\R \times S_{\tau}^1$. 
We focus on the distributions and expectation values of the Polyakov loop, 
the dependence of the argument of the Polyakov loop on $L_{s}$. 
In addition, we also investigate the ``pseudo-entropy" density $\propto\langle T_{xx}-T_{\tau\tau}\rangle$,
which is the counterpart of the thermal entropy density in the $\Z_{N}$-TBC case.
For small inverse coupling $\beta$ (or large $L_{\tau}$), 
the value of the Polyakov loop for each configuration 
concentrates around the origin on the complex plane and 
its expectation value is zero ($|\langle P\rangle| = 0$) 
as a result of the exact $\Z_N$ symmetry. 
The distribution of the Polyakov loop spread around the origin 
broadly and isotropically as $\beta$ increases. 
We find that, in high $\beta$ regions (or small $L_{\tau}$ regions), the distribution forms a regular $N$-sided-polygon shape and the expectation value of the Polyakov loop is still consistent with zero ($|\langle P\rangle|\sim 0$).
By studying the $L_{s}$-dependence of the Polyakov loop, 
we also show the existence of fractional instantons and bions, which cause tunneling transition 
among the classical $N$ vacua, 
leading to the stabilization of the $\Z_N$-symmetric vacuum.
For much higher $\beta$, the statistics in our simulation are less than the auto-correlation time for $|\langle P\rangle|$ and 
the results of the simulation gets unreliable.
Even in such high $\beta$ regions, we find that a regular $N$-sided-polygon shape of the Polyakov-loop distribution tends to be restored and $|\langle P\rangle|$ gets smaller by increasing the number of samples.
These results indicate the stability 
of the ${\mathbb Z}_{N}$ symmetry and also imply that 
the seemingly broken ${\mathbb Z}_{N}$ symmetry 
at extremely high $\beta$ can be an artifact 
due to insufficient statistics.
In order to study the adiabatic continuity from another viewpoint, 
we calculate the $\beta$ dependence of 
the pseudo-entropy density, 
$s \propto \langle T_{xx}-T_{\tau \tau} \rangle$, 
and find no phase transition 
as $\beta$ increases from small to large values.
Furthermore, our result indicates that 
the pseudo-entropy density vanishes in the large-$N$ limit.
It suggests the volume independence 
in the whole $\beta$ regime in the large-$N$ limit, 
consistent with the analytical study of 
``the large-$N$ volume independence" 
in Ref.~\cite{Sulejmanpasic:2016llc}.

This paper is organized as follows.
In Sec.~\ref{sec:CP}, we introduce the model 
in the continuum limit and review its properties.
In Sec~\ref{sec:setup}, we review the results of lattice simulation for the model with the PBC.
In Sec.~\ref{sec:ZN}, we show the results of the lattice Monte Carlo simulation for the model with the $\Z_N$ TBC.
In Sec.~\ref{sec:entropy}, we measure the pseudo-entropy density and discuss its implication on the adiabatic continuity.
Section~\ref{sec:summary} is devoted to a summary and discussion.


\section{Two-dimensional ${\mathbb C}P^{N-1}$ model}
\label{sec:CP}

In this paper, we investigate ${\mathbb C}P^{N-1} = 
{\rm SU}(N)/({\rm SU}(N-1)\times {\rm U}(1))$ sigma models 
(without the topological $\theta$-term) on $\R\times S^{1}$, 
with attentions to the $\Z_N$-TBC, $1/N$ fractional instantons and $\Z_{N}$ intertwined symmetry.
In this section we review these notions.

\subsection{Basics of the model}

Let $\omega(x)$ be an $N$-component vector of 
complex scalar fields, and $\phi(x)$ be 
a normalized complex $N$-component vector 
composed of $\omega$ as 
$\phi(x) \equiv \omega(x)/\sqrt{\omega^\dagger \omega}$. 
The action of the ${\mathbb C}P^{N-1}$ model 
in Euclidean two dimensions
is given by 
\begin{align}
S&={1\over{g_{0}^{2}}}\int d^{2}x \, (D_{\mu}\phi)^{\dag} (D_{\mu} \phi). 
\end{align}
where $g_{0}$ is the bare coupling constant,
$d^{2} x\equiv dxd\tau$ and 
the indices $\mu,\nu=1,2$ label the $x,\tau$ directions.
The covariant derivative is defined as 
$D_{\mu}= \partial_{\mu}+iA_{\mu}$ 
with a composite gauge field $A_{\mu}(x) \equiv i\phi^{\dag}{\overleftrightarrow\partial}_{\mu}\phi$.
It is notable that global ``flavor" symmetry of the model is PSU($N$)=SU($N$)/${\mathbb Z}_{N}$, 
where the ${\mathbb Z}_{N}$ center is removed 
since it coincides with a subgroup of 
the U($1$) gauge symmetry and hence is redundant.
All through this paper, we consider (or approximate) the model on $\R\times S^{1}$,
and regard $x$ and $\tau$ as coordinates of uncompactified (large) and compactified (small) directions, respectively.

This model has instanton solutions characterized by 
the topological charge representing 
$\pi_2({\mathbb C}P^{N-1}) \simeq {\mathbb Z}$ 
\begin{align}
Q&={1\over{2\pi}}\int d^{2}x \; i\epsilon_{\mu\nu} (D_{\nu}\phi)^{\dag} (D_{\mu} \phi)={1\over{2\pi}}\int d^{2}x \epsilon_{\mu\nu}\partial_{\mu} A_{\nu}\,.
\label{Qdef}
\end{align}

The simplest case, or the ${\mathbb C}P^1$ model, 
is equivalent to the O($3$) nonlinear sigma model, 
thus it can be also described by three real scalar fields 
${\bf m}(x)=(m^1(x),m^2(x),m^3(x))$ 
with a constraint ${\bf m}(x)^2=1$.
Its action is given by
\begin{align}
S = \frac{1}{g_{0}^2}
\int d^2x \, \partial_{\mu} {\bf m} \cdot  \partial_{\mu} {\bf m} ,
\end{align}
where the relation between the real scalar field ${\bf m}(x)$ and the complex two-component complex field $\omega(x)$ is
\begin{eqnarray}
{\bf m}(x)= \phi^{\dag} (x) \vec{\sigma} \phi(x) 
= \left(\frac{\omega^{*1}\omega^2 + \omega^{*2}\omega^1}{\omega^\dagger(x) \omega(x)} \,,\,
-i\frac{\omega^{*1}\omega^2 - \omega^{*2}\omega^1}{\omega^\dagger(x) \omega(x)} \,,\,
\frac{|\omega^1|^2 - |\omega^2|^2}{\omega^\dagger(x) \omega(x)} \right) \,,
\label{eq:three-vector}
\end{eqnarray}
with the Pauli matrices $\vec{\sigma}$. 
In this description, the configuration $\phi = (1,0)^{T}$ corresponds to the north pole
${\bf m}=(0,0,1)$, while $\phi = (0,1)^{T}$ corresponds to the south pole ${\bf m}=(0,0,-1)$.


\subsection{${\mathbb Z}_{N}$-twisted boundary condition and fractional instantons}

The boundary condition we mainly focus on 
is the $\Z_N$-symmetric twisted boundary condition 
($Z_N$ TBC).
The ${\mathbb Z}_{N}$-TBC in a compactified direction 
can be expressed as
\begin{equation}
\phi(x, \tau+L_\tau) = \Omega \,\phi(x,\tau)\,,
\,\,\,\,\,\,\,\,\,\,\,\,\,
\Omega ={\rm diag.}\left[1, e^{2\pi i/N}, e^{4\pi i/N},\cdot\cdot\cdot, e^{2(N-1)\pi i/N} \right]\,,
\label{ZNC}
\end{equation}
where $L_{\tau}$ is the compactification circumference. 
We note that this ${\mathbb Z}_{N}$-TBC is equivalent to the existence of the following Wilson-loop holonomy of the background SU($N$) gauge field in the compactified direction:
\begin{equation}
\langle A_{\tau} \rangle = \diag.  (0, 2\pi/N, \cdot\cdot\cdot, 2(N-1)\pi/N)\,.
\label{WHN}
\end{equation}
Let us discuss the symmetry of the present model.
Consider the $\Z_N$ subgroup of the flavor SU($N$) transformation generated by 
\beq
\phi=(\phi_{1},\phi_{2},...,\phi_{N})^{T}
\,\mapsto\, S\phi=(\phi_{N},\phi_{1},...\phi_{N-1})^{T}, 
\quad\quad
S=\left({\renewcommand{\arraystretch}{0.6}
{\arraycolsep 1.1mm\begin{array}{ccccc}
0&0&\cdots&0&1\\
1&0&\cdots&0&0\\
0&1&\cdots&0&0\\
 \vdots&\vdots & \ddots &\vdots &\vdots\\
0&0&\cdots&1&0\\
\end{array}}}\right). 
\eeq
This transformation alone does not keep the $\Z_N$-TBC 
but becomes a symmetry which 
does not change the boundary condition 
if it is combined with the large gauge transformation
\beq
\phi \rightarrow e^{\frac{2\pi i \tau}{NL_\tau}} \phi. 
\eeq 
Explicitly, we can confirm that 
the intertwined transformation 
\beq
\phi \rightarrow \phi' = e^{-\frac{2\pi i \tau}{NL_\tau}} S \phi
\label{eq:intertwined}
\eeq
does not change the boundary condition as
\beq
\phi'(x,\tau+L_{\tau})= e^{\frac{2\pi i (\tau+L_\tau)}{NL_\tau}} S \Omega \phi(x,\tau) = e^{\frac{2\pi i L_\tau}{NL_\tau}} S\Omega S^{-1} \phi'(x,\tau) = \Omega \phi'(x,\tau),
\label{eq:shift}
\eeq 
where we have used the fact that 
the matrix $\Omega$ is invariant 
under the combined transformation
\beq
\Omega ~\rightarrow~ e^{\frac{2\pi i \tau}{NL_\tau}} S\Omega S^{-1}  = \Omega.
\eeq 
Therefore thus model has the intertwined $\Z_N$ symmetry, which is the invariance under the simultaneous $\Z_{N}$ large gauge transformation and $\Z_{N}$ flavor-shift transformations.
This is a significant difference of the model with the $\Z_N$ TBC from the model with the PBC.
We also note that the continuous part of the flavor symmetry $SU(N)/ \Z_N$ is explicitly broken to $U(1)^{N-1}/\mathbb{Z}_N$.

The existence of the exact $\Z_N$ symmetry 
indicates that the model has $N$ vacua at the classical level.
This fact becomes clear by looking into 
the dimensionally reduced effective one-dimensional model 
under the $S^{1}$-compactification with the $\Z_N$ twist: 
in the effective theory, the twist gives 
a potential with $N$ vacua related by the $\Z_N$ symmetry. 
For the $\C P^{1}$ model ($N=2$),
the two equivalent vacua are located 
at the north pole $\phi = (1,0)^{T}$ (${\bf m}=(0,0,1)$) 
and the south pole  $\phi = (0,1)^{T}$ (${\bf m}=(0,0,-1)$).

We now introduce fractional instantons \cite{Eto:2004rz,Eto:2006mz,Eto:2006pg}
in the ${\mathbb C}P^{1}$ model 
satisfying the ${\mathbb Z}_{2}$-TBC, 
which is given by 
\begin{eqnarray}
&& \omega (x,\tau+L_{\tau}
) = {\rm diag.}[1,e^{\pi i}] \omega (x,\tau)
= {\rm diag.}[1,-1] \omega (x,\tau),  
\end{eqnarray}
or equivalently
\begin{eqnarray}
&& (m^1(x,\tau+L_{\tau}),m^2(x,\tau+L_{\tau}),m^3(x,\tau+L_{\tau})) 
= (-m^1(x,\tau),-m^2(x,\tau),m^3(x,\tau))\,.
\end{eqnarray}
By defining the complex coordinate $z = x+i\tau$,
we can express one of the fractional instanton solutions as
\begin{eqnarray}
&& \omega
 = \left(1,a\,e^{+\pi z} \right)^{T}\,, \quad
 \label{eq:fractional}
\end{eqnarray}
where two real moduli parameters (a position and phase) 
are combined into a complex constant $a$.
This configuration is a Bogomol'nyi-Prasad-Sommerfield (BPS) 
solution, which is holomorphic and depends only on $z$.
We also have other fractional solutions including
the other BPS fractional instanton $ \omega = \left(1,a\,e^{-\pi z} \right)^{T}$,
two anti-BPS fractional instantons $\omega = \left(1,a\, e^{+\pi \bar z} \right)^{T}$
and $\omega = \left(1,a\,e^{-\pi \bar z} \right)^{T}$.

The fractional instanton configuration becomes $\phi = (1,0)^{T}$ 
(${\bf m}=(0,0,1)$) at $x \to -\infty$ while it becomes $\phi = (0,1)^{T}$ 
(${\bf m}=(0,0,-1)$) at $x \to +\infty$.
It is notable that, at each constant $\tau$ slice, it corresponds to a path connecting 
the north pole ${\bf m}=(0,0,+1)$ and the south pole ${\bf m}=(0,0,-1)$ in the target space.
The phase of $a$ determines the way 
how the vector ${\bf m}=(0,0,+1)$ changes to ${\bf m}=(0,0,-1)$
and can be interpreted as a U($1$) modulus 
localized on the domain-wall at 
$x_{0}={1\over{\pi}}\log{1\over{|a|}}$.
Under the $\Z_2$-TBC, 
the $U(1)$ modulus is twisted by $\Z_2$ along the domain wall. 
It means that, when the constant $\tau$ slice is changed 
from $\tau=0$ to $\tau=L_{\tau}$, 
the path in the target space sweeps 
a half of the sphere of the target space.
Therefore, the configuration is a map from 
the space ${\mathbb R} \times S^1$ 
to a half of the target space.
This is the reason why it carries 
a half of the unit instanton charge $Q= 1/2$
while the anti-BPS configurations carry $Q=- 1/2$.
One finds that the topological charge density has 
no dependence on the compactified direction $\tau$.


The fractional instantons 
in the ${\mathbb C}P^{N-1}$ model are 
classified in a parallel manner. 
The configuration (\ref{eq:fractional}) of the 
${\mathbb C}P^{1}$ model can be generalized 
to the $N$-vector $\omega$ for the ${\mathbb C}P^{N-1}$ model 
with the ${\mathbb Z}_N$-TBC (\ref{ZNC}) as
\begin{eqnarray}
\omega  = \left(0\,,\, \cdots\,,\, 0\,,\,1\,,\,a\, e^{+2\pi z/N}\,,\, 0\,,\,\cdots \right)^{T}\,.
\end{eqnarray}
These configurations carry $1/N$ unit of the instanton charge 
($Q= 1/N$).
These fractional instantons correspond to 
the tunneling transitions (domain walls) 
among $N$ classical vacua.
It means that they stabilize the $\Z_N$-symmetric vacua 
at the quantum level.
In the quantum-mechanical limit with $L_{\tau}\to 0$, 
we can easily show that the $\Z_{N}$ symmetry is preserved
at the quantum level due to the fractional instantons.
The conjecture on the adiabatic continuity of 
the vacuum structure states~\cite{Sulejmanpasic:2016llc} that ``{\it the $\Z_N$-symmetric vacua continue to be unbroken from the small $S^{1}$ regime to the large $S^{1}$ regime}".
One of the purposes of the present work is to study the conjecture in terms of lattice Monte Carlo simulation.


\section{Review of simulation for periodic boundary condition}
\label{sec:setup}


\subsection{Lattice setup}
We now review the results for PBC, in particular the expectation value of the Polyakov loop \cite{Fujimori:2019skd}.
The lattice action of the two-dimensional 
$\C P^{N-1}$ sigma model \cite{Berg:1981er,Campostrini:1992ar,Farchioni:1993jd,Alles:2000sc,Flynn:2015uma,Abe:2018loi,Fujimori:2019skd} is
\begin{equation}
S=N\beta \sum _{n,\mu} \left( 2- \bar{\phi}_{n+\hat{\mu}}\cdot \phi_{n} \,
\lambda_{n,\mu} - \bar{\phi}_{n}\cdot \phi_{n+\hat{\mu}} 
\bar{\lambda}_{n,\mu} \right)\,,
\label{eq:latt-action}
\end{equation}
with $\bar{\phi}_{n} \cdot \phi_{n} =1$ and $\lambda_{n,\mu}$ 
being a link variable corresponding to the auxiliary U($1$) gauge field. 
The lattice sites $n=(n_x, n_\tau)$ run as $n_x = 1, \cdots, N_s$ and $n_\tau = 1, \cdots, N_\tau$
and $N\beta$ is equivalent to the inverse of the bare coupling $1/ g_{0}^{2}$.
The circumferences of $S_s^{1}$ and $S^1_\tau$ are given 
in terms of the lattice spacing $a$ as $L_{s} = N_{s} a$ and $L_{\tau} = N_{\tau} a$, respectively.

The renormalization group tells us the relation 
between $\beta$ and the lattice spacing $a$
\beq
\Lambda_{\overline{MS}} \,a = (2\pi\beta)^{{2\over{N}}} e^{-2\pi\beta}\,,
\eeq
where the renormalized coupling in the $\overline{MS}$ scheme diverges at $\Lambda_{\overline{MS}}$. 
By comparing $\Lambda_{\overline{MS}}$ \cite{Campostrini:1992ar} 
and the lattice $\Lambda$ scale $\Lambda_{lat}$ for Eq.(\ref{eq:latt-action}), 
we obtain the following relation,
\begin{equation}
\Lambda_{lat} \, a = {1 \over{\sqrt{32}}} (2\pi\beta)^{2\over{N}} 
e^{-2\pi\beta -{\pi\over{2N}}}\,,
\label{eq:beta-a-relation}
\end{equation}
which relates $a$ with $\beta$ for each $N$, where $\Lambda_{lat}$ works as a reference scale. 
The parameter set with $L_s \gg L_\tau$ was taken to approximately simulate the model on $\mathbb R \times S^1$, where $L_{\tau}$ corresponds to an inverse temperature $1/T$ and smaller $L_{\tau}$ or higher $\beta$ with fixed $N_\tau$ corresponds to higher $T$.


\subsection{Lattice simulation for periodic boundary condition}

The Polyakov loop on the lattice is given by
\begin{equation}
P\equiv {1\over{N_{s}}} \sum_{n_x}\prod_{n_\tau} \lambda_{n,\tau}.
\end{equation}
It is an order parameter of the ${\mathbb Z}_{N}$ symmetry, whose transformation is given by
\begin{equation}
P \,\mapsto\, e^{2\pi i/N} P\,.
\end{equation} 
While this is an exact symmetry and the expectation value of the Polyakov loop $\langle P \rangle$ is its order parameter 
in the model with $\Z_{N}$-TBC, 
the model with the PBC  
is not $\Z_N$-symmetric.
However, $\langle P \rangle$ is still of physical importance
in the PBC case.
Based on the relation between the Polyakov and 
Wilson loops via the clustering property, 
one finds that the expectation value of the Polyakov loop vanishes
($\langle P \rangle=0$) in the confinement phase with a nonzero string tension in the $L_\tau \ll L_s$ system.
Thus, $\langle P \rangle$ can be adopted as the order parameter of the confinement-deconfinement transition in the system.
Below, we summarize the main results of the lattice simulation 
for the model with PBC.
All these results strongly indicate a crossover behavior between the confinement and deconfinement phases.

\begin{enumerate}

\item
The distribution plots of the Polyakov loop for $N=3,5,10,20$ exhibit 
that the Polyakov loops are distributed around the origin at low $\beta$ and gradually get away from the origin at higher $\beta$. 

\item
The absolute values of the expectation values of the Polyakov loop 
$|\langle P \rangle|$ as functions of $\beta$ for 
$N=3,5,10,20$ and ($N_s,N_\tau$) $=$ ($200,8$) indicate 
$|\langle P \rangle| \approx 0$ for low $\beta$ (large $L_{\tau}$). 
The value of $|\langle P \rangle| $ is independent of 
$\beta$ in $\beta \ll \beta_c$ since the IR scale of the system is 
the confinement scale, but not $1/L_{\tau}$, in the confining phase.
We here denote $\beta_c$ as a pseudo-critical $\beta$, which is defined
from the peak position of $\chi_{\langle | P| \rangle} = V (\langle |P|^2 \rangle - \langle |P| \rangle^2)$.
As $\beta$ increases across $\beta_c$, $|\langle P \rangle|$ gradually increases and we find 
$|\langle P \rangle| \not= 0$ for high $\beta$ (small $L_{\tau}$).

\item
The Polyakov-loop susceptibility $\chi_{\langle | P| \rangle}$ as a function of $1/L_\tau$ shows that the peak is broad for small $N$ but gets sharper as $N$ increases.

\item
The volume dependence of the peak value of $\chi_{\langle | P| \rangle}$ is measured by simulating $N_{s}=40,80,120,160,200$ with $N_{\tau}=8$ fixed.  
The fit of data points for $N_s = 80,120,160,200$ 
by a function $\chi_{\langle|P|\rangle,{\rm max}} \,=\, 
a + c N_{s}^{p}\,$ \cite{Fukugita:1990vu} indicates that the 
best fit values of the exponent are $p<1$ for $N=3,5,10,20$, 
in conformity with a crossover behavior.

\end{enumerate}


\section{Simulation results for ${\mathbb Z}_{N}$-twisted boundary condition}
\label{sec:ZN}


In the ${\mathbb C}P^{N-1}$ model on $\R\times S^{1}$ 
with the ${\mathbb Z}_{N}$-TBC,
the ${\mathbb Z}_{N}$ intertwined symmetry \eqref{eq:intertwined} is exact. 
This ${\mathbb Z}_{N}$ symmetry is conjectured to be unbroken even at high temperature (small $L_{\tau}$) due to transitions among the ${\mathbb Z}_{N}$ vacua via the fractional instantons \cite{Dunne:2012ae,Dunne:2012zk}.

We here perform the Monte Carlo simulation with 
the ${\mathbb Z}_{N}$-TBC imposed 
on the small compactified direction 
$S_{\tau}^{1}$ for $N=3,5,10,20$ and 
$(N_{s}, N_{\tau})=(200,8), \; (400,12)$.
The simulation setup except for the boundary condition 
is the same as the previous work~\cite{Fujimori:2019skd}.
In this section, we show the results of distribution plots and expectation values of the Polyakov loop operator for these cases, which indicate the stability of the $\Z_{N}$-symmetric vacua in the model with the $\Z_N$-TBC.
We also find configurations corresponding to 
fractional instantons and bions, 
which work to stabilize the ${\mathbb Z}_{N}$ symmetry.

We note that there are few lattice studies on the ${\mathbb C}P^{N-1}$ model with the ${\mathbb Z}_{N}$-TBC although the lattice simulations of Yang-Mills theory and QCD with the twisted boundary condition have been well performed \cite{Iritani:2015ara,Misumi:2015hfa,Itou:2018wkm}.

\begin{figure}[t]
\includegraphics[width=0.99\linewidth]{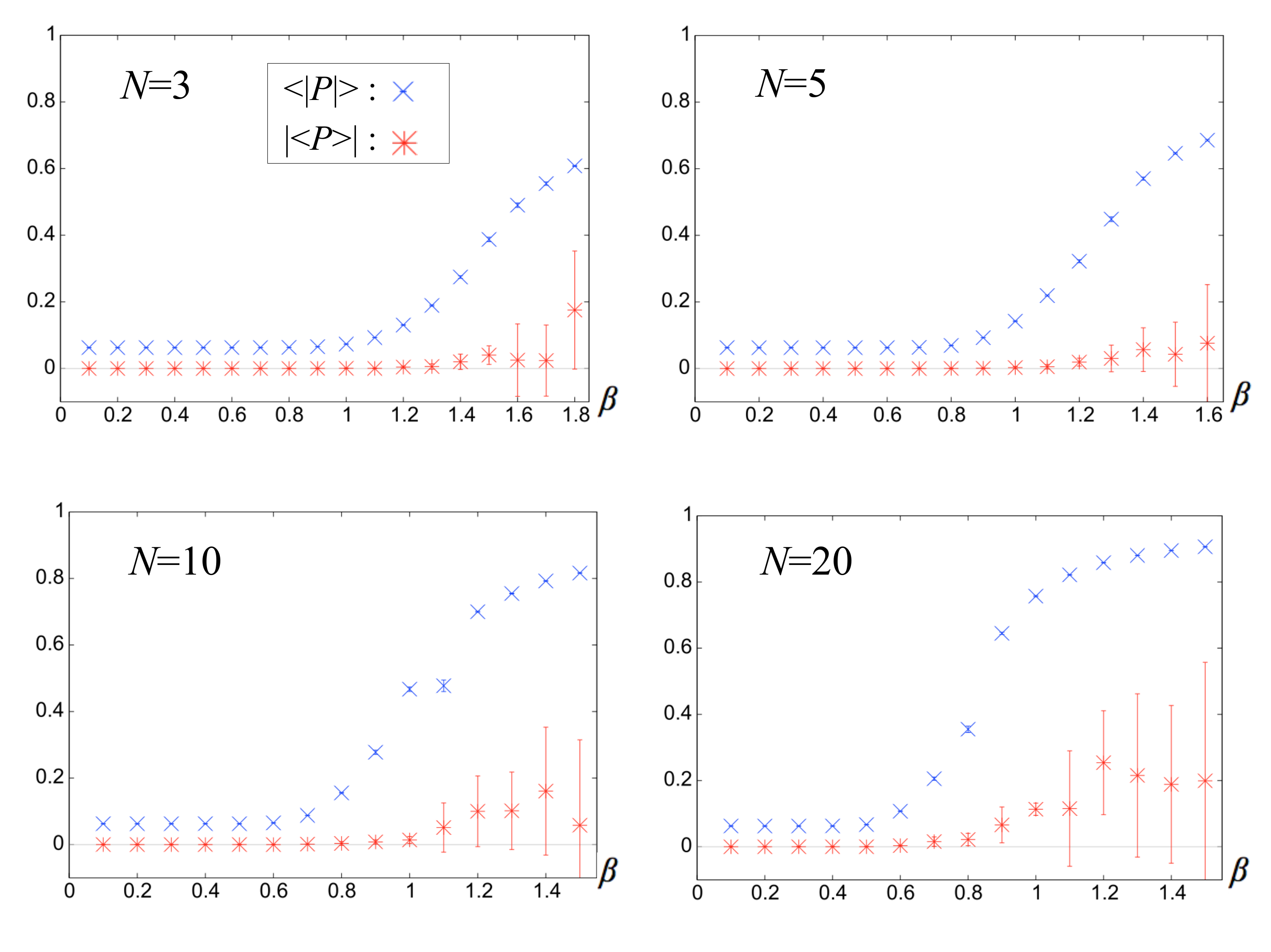}
 \caption{
Expectation values of the Polyakov loop for $N=3,5,10,20$ with $(N_{s},N_{\tau})=(200,8)$ with ${\mathbb Z}_{N}$-TBC.
Red points indicate $|\langle P \rangle|$ while blue points $\langle |P|\rangle$. 
For low $\beta$, $|\langle P \rangle|$ (red points) is consistent with zero. For high $\beta$, they are still consistent to zero although $\langle |P|\rangle$ gets large.
}
\label{fig:Ploop_ZN}
\end{figure}

\begin{figure}[t]
 \includegraphics[width=0.4\linewidth]{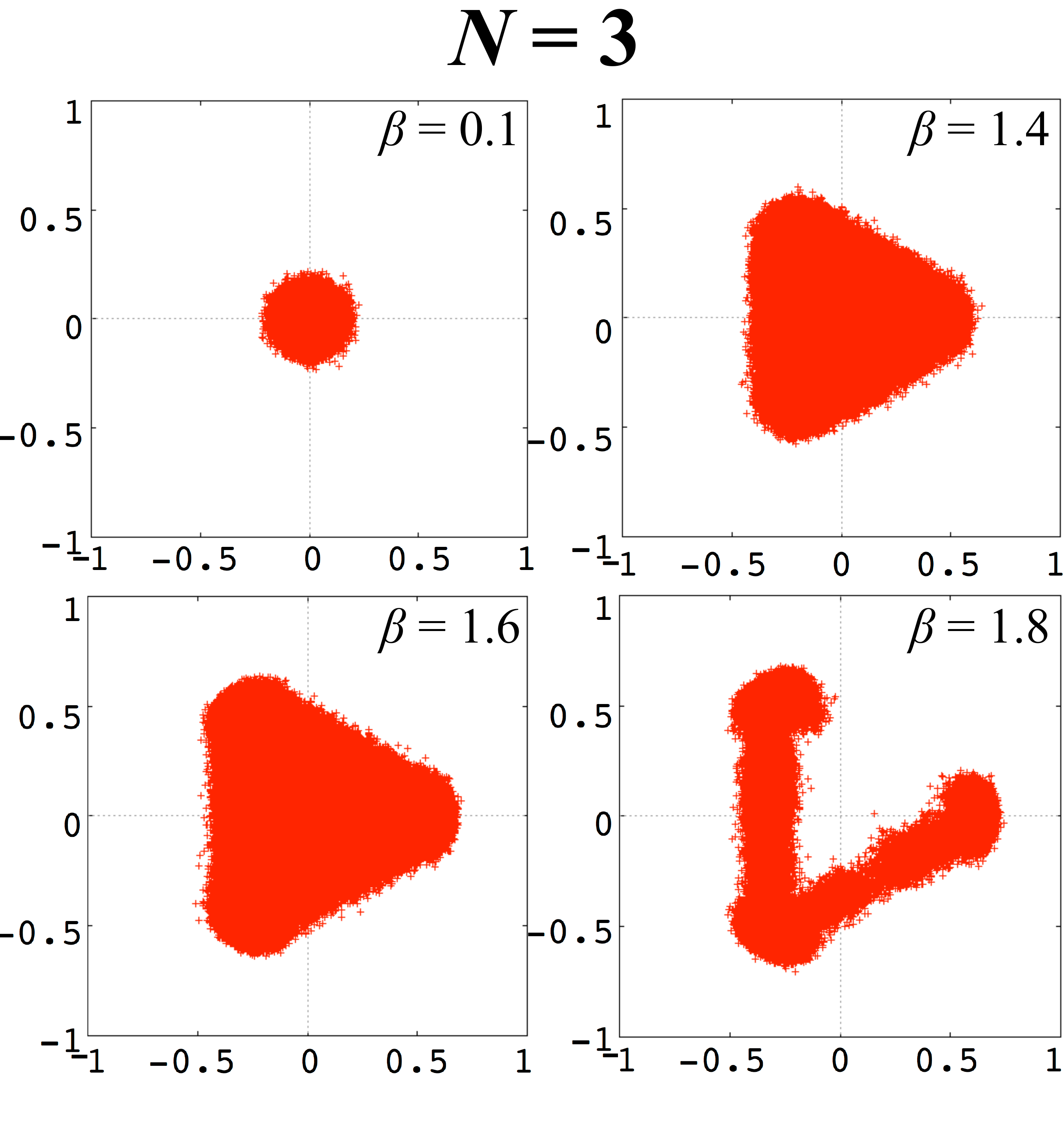}
  \includegraphics[width=0.4\linewidth]{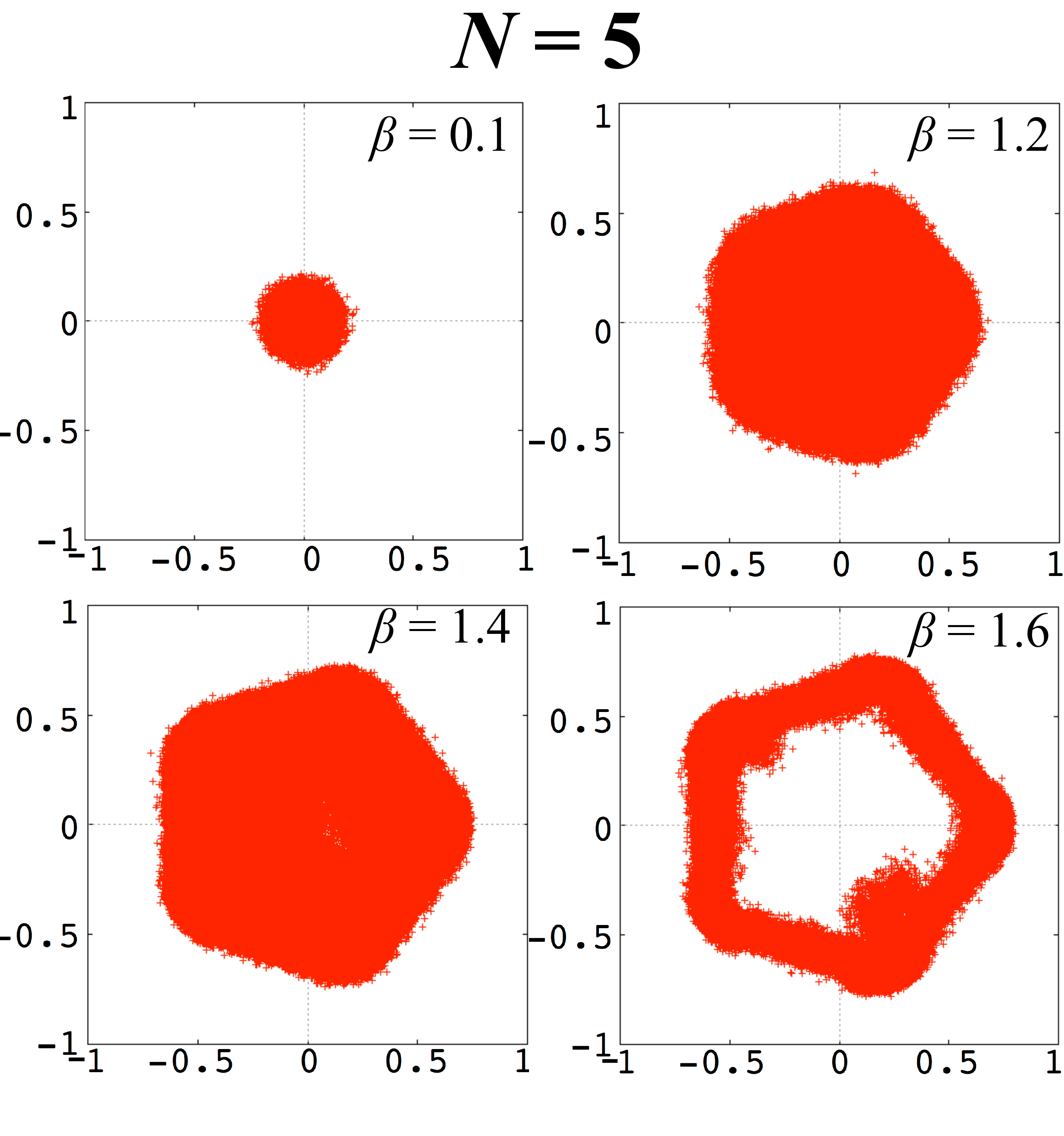}
 \includegraphics[width=0.4\linewidth]{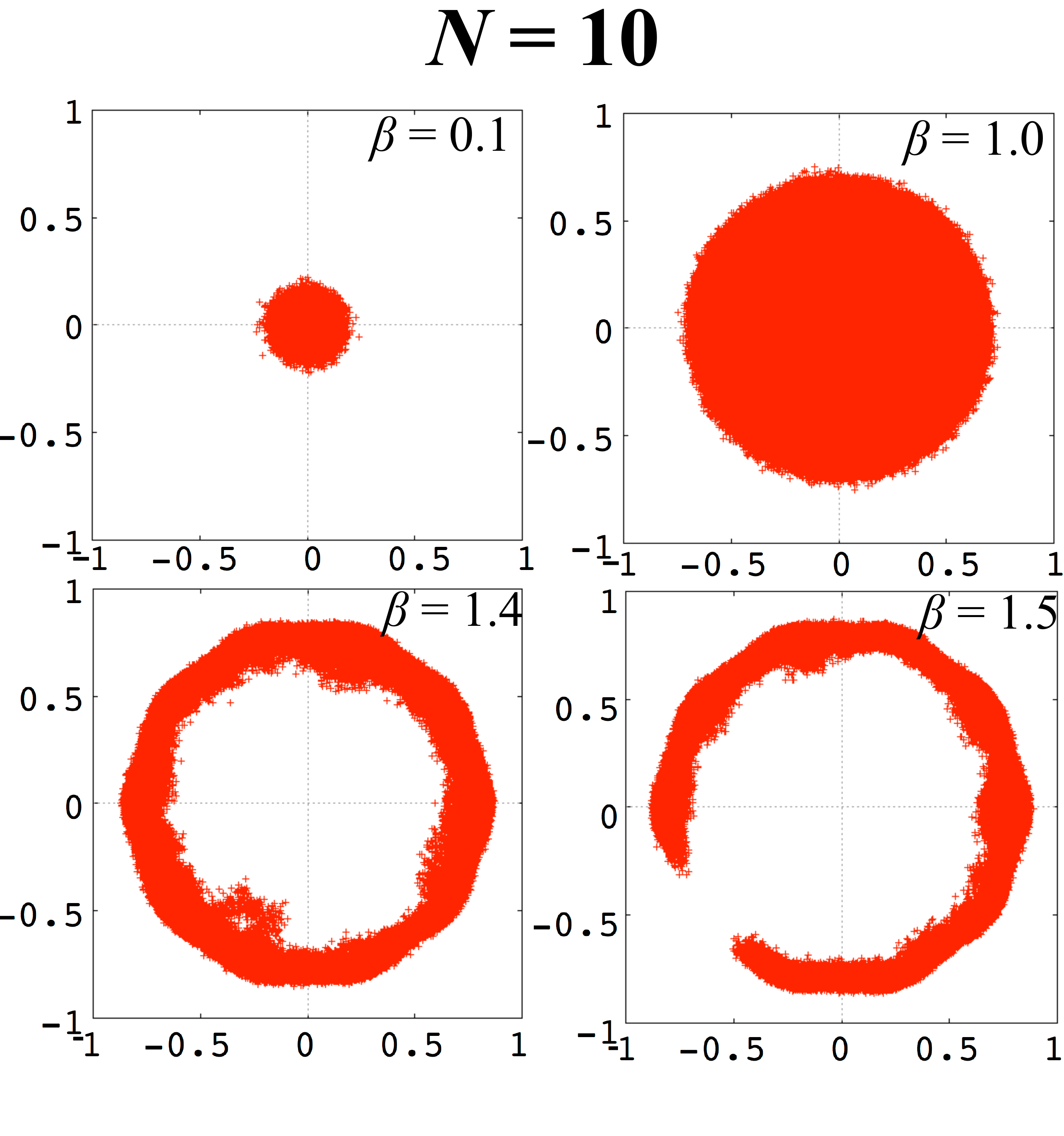}
   \includegraphics[width=0.4\linewidth]{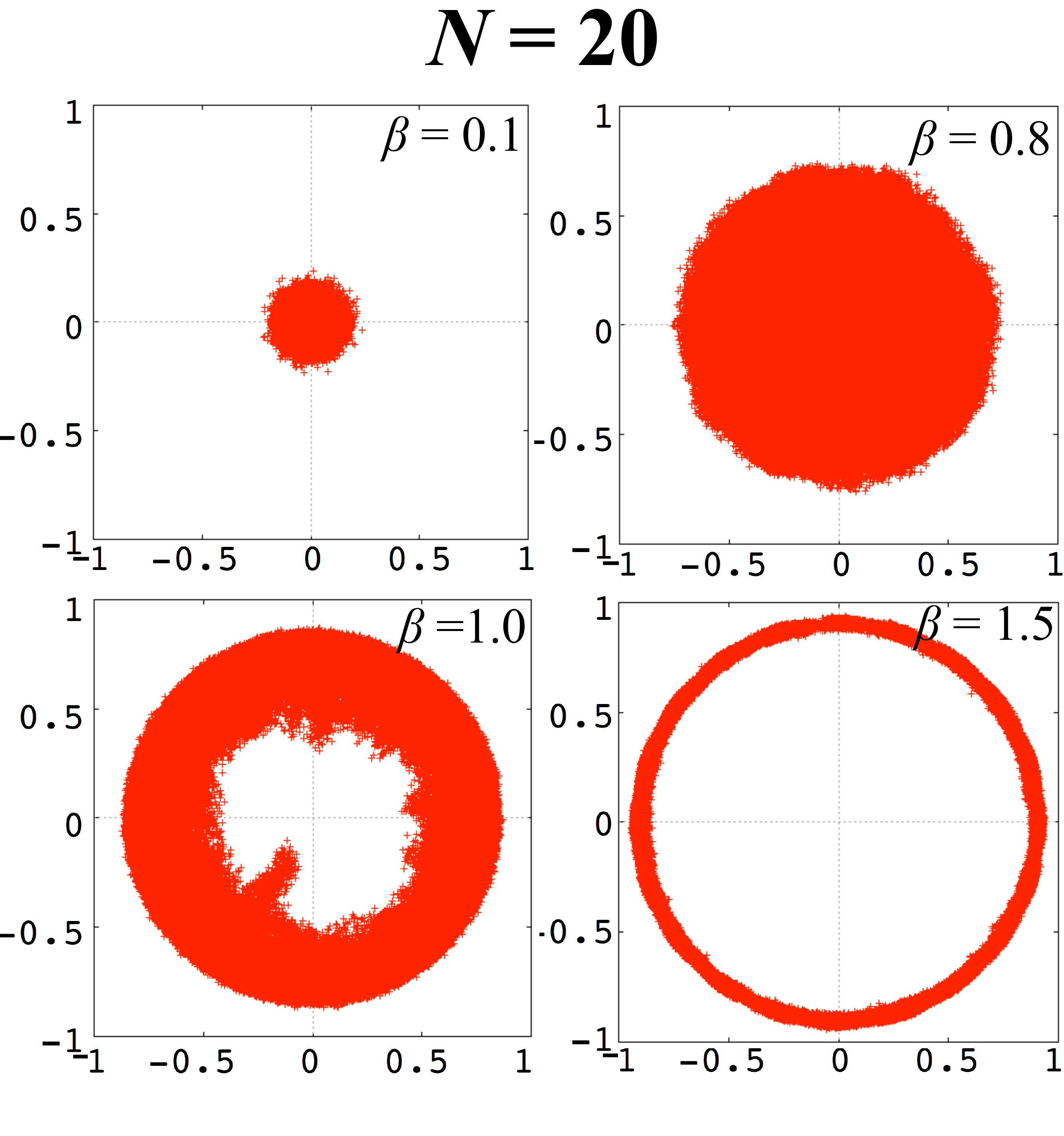}
 \caption{Distribution plots of the Polyakov loop for $N=3,5,10,20$ with $(N_{s},N_{\tau})=(200,8)$ with ${\mathbb Z}_{N}$-TBC. Regular $N$-sided polygon shapes appear at high $\beta$.}
\label{fig:dist_ZN}
\end{figure}

\subsection{Polyakov-loop expectation values and distribution plots}\label{sec:Ploop}

We first show the $\beta$ dependence of the expectation values of the Polyakov-loop operator.
We note that a larger $\beta$ corresponds to a smaller $L_{\tau}$.
We now focus on two types of the expectation values, one of which is the absolute values of the expectation values of the Polyakov loop, 
$|\langle P \rangle|$, and the other of which is the expectation values of the absolute values of the Polyakov loop, $\langle |P|\rangle$.
The former, $|\langle P \rangle|$, is a genuine order parameter of the ${\mathbb Z}_{N}$ symmetry, while the latter, 
$\langle |P| \rangle$, is not an order parameter of ${\mathbb Z}_{N}$ symmetry, 
indicating the averaged distance of the Polyakov-loop values from the origin in the complex plane. 
For the case of the PBC, these two quantities 
show a crossover behavior and have an identical behavior except for a small $\beta$ region, 
where we have $|\langle P \rangle|=0$ 
while $\langle |P| \rangle$ takes a 
finite value due to the finite volume effect \cite{Fujimori:2019skd}. 
We show below that it is not the case with the $\Z_N$-TBC.

We now write down the procedure for generating configurations in order to investigate the $\beta$ dependence of the distribution and the expectation values of the Polyakov loop:
(1) We start to generate configurations at low $\beta$ such as $\beta=0.1$. (2) We then pick up one of the configurations and adopt it as an initial configuration to generate them at $0.1$ higher $\beta$. (3) We repeat this procedure and generate configurations from low to high $\beta$.

In our simulations the maximal numbers of Monte Carlo steps are $802,000$ for $N=3,5$ and $602,000$ for $N=10,20$, where we throw away the first $2,000$ steps to thermalize the systems.
Therefore, the numbers of samples are $N_{\rm sweep} = 800,000$ for $N=3,5$ and $N_{\rm sweep} = 600,000$ for $N=10,20$.
Hereafter we denote the number of samples or sweeps as $N_{\rm sweep}$.
In Fig.~\ref{fig:Ploop_ZN}, we depict $|\langle P \rangle|$ as red points and $\langle |P| \rangle$ by blue points for the $\mathbb Z_{N}$-TBC, $N=3,5,10,20$ with $(N_{s}, N_{\tau})=(200,8)$.
We carry out the jackknife error estimation with the binning method and take the bin size $N_{\rm bin}$ to be larger than the auto-correlation time.
We also show the distribution plots of the Polyakov loop for these cases with several different $\beta$ in Fig.~\ref{fig:dist_ZN}.

In Fig.~\ref{fig:Ploop_ZN}, $\langle |P|\rangle$ (blue points) gradually increases in $\beta_c < \beta$ for each $N$, {\it e.g.} $\beta_c \approx 1.0$ for $N=3$ and $\beta_c \approx 0.5$ for $N=20$, since the distribution of the Polyakov loop gradually spreads as shown in Fig.~\ref{fig:dist_ZN}.
We here define $\beta_{c}$ as the value of $\beta$ at which $\langle |P|\rangle$ starts to increase
and $1/L_{\tau}$ becomes the IR scale of the system.
In $\beta < \beta_c $ regime, the values of $\langle |P|\rangle$ is independent of $\beta$ and take the same values as those for PBC as we will show in Fig.~\ref{fig:Ploopab_ZNvsPBC}. 
It indicates that the IR scale of the system changes from $\Lambda_{\C P^{N-1}}$ to $1/L_{\tau}$ at $\beta_{c}$.

On the other hand, the genuine $\Z_N$ order parameter $|\langle P \rangle|$ (red points) is almost consistent to zero within statistical errors even at a high $\beta$ region ($\beta>\beta_{c}$). 
This behavior originates in the fact that the distribution of the Polyakov loop spreads isotropically around the origin.
Let us investigate this $\Z_N$ order parameter $|\langle P \rangle|$ in more detail by comparing it to the distribution plot in Fig.~\ref{fig:dist_ZN}.
For low $\beta$, $|\langle P \rangle|$ is zero with quite small errors and the distribution of the Polyakov loop is concentrated around the origin in the complex plane. This behavior reflects the exact ${\mathbb Z}_{N}$ symmetry of the present model.
For intermediate $\beta$, the distributions of the Polyakov loop are 
spreading out and form regular $N$-sided polygon shapes in the complex plane. 
The value of $|\langle P \rangle|$ is consistent with zero but the errors get larger.
For high $\beta$, the distributions still form regular $N$-sided polygon shapes, but they tend to assemble around the $N$ sides of the polygons. At this high $\beta$ ($\beta>\beta_{c}$), $|\langle P \rangle|$ is still consistent with zero or takes a very small value with large errors due to the broadly and isotropically spread distributions.
These results clearly indicate the stability of the $\Z_N$ symmetry in the model.
As we will see in the next subsection, this stability of the ${\mathbb Z}_{N}$ symmetry originates in the tunneling transition among the equivalent $N$ classical vacua via fractional instantons.

\begin{figure}[t]
 \includegraphics[width=0.99\linewidth]{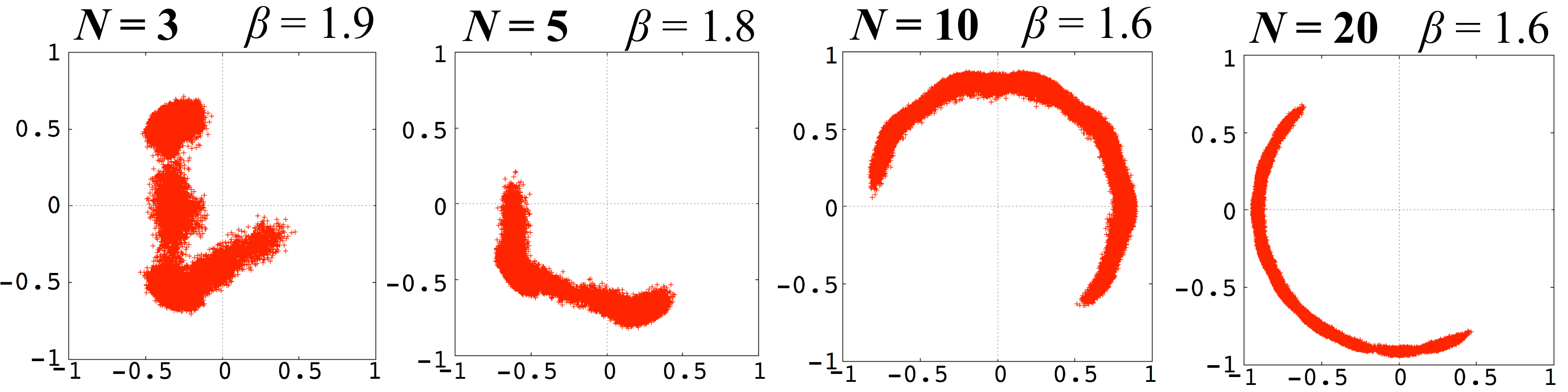}
 \caption{Distribution plots of the Polyakov loop for $N=3,5,10,20$ with $(N_{s},N_{\tau})=(200,8)$ with ${\mathbb Z}_{N}$-TBC lose their $N$-sided-polygon shapes at quite high $\beta$.}
\label{fig:dist_ZN_h}
\end{figure}

We now discuss the validity of our simulation results.
As will be discussed in App.~\ref{sec:bin-size}, we have estimated the auto-correlation times for $|\langle P \rangle|$ by studying the bin-size dependence of the Jackknife error bars.
Based on the estimated auto-correlation time, we consider that our simulations, whose main results are shown in Figs.~\ref{fig:Ploop_ZN} and \ref{fig:dist_ZN}, are performed with the larger number of samples than the auto-correlation times.
On the other hand, for much higher $\beta$, the auto-correlation times seem to be larger than the number of samples ($800,000$ for $N=3,5$ and $600,000$ for $N=10,20$), and thus we cannot obtain meaningful results of the Polyakov-loop expectation values in the $\beta$ region so far. 
Indeed, in this $\beta$ region, the regular $N$-sided polygon shapes of the Polyakov-loop distribution lose
their shapes as shown in Fig.~\ref{fig:dist_ZN_h}.
We consider that this behavior is just an artifact due to the shortage of statistics since the regular $N$-sided polygon shapes of the Polyakov-loop distribution are getting restored as we adopt a larger number of samples even for these cases as we will discuss in the next-next subsection.

\begin{figure}[t]
 \includegraphics[width=1.0\linewidth]{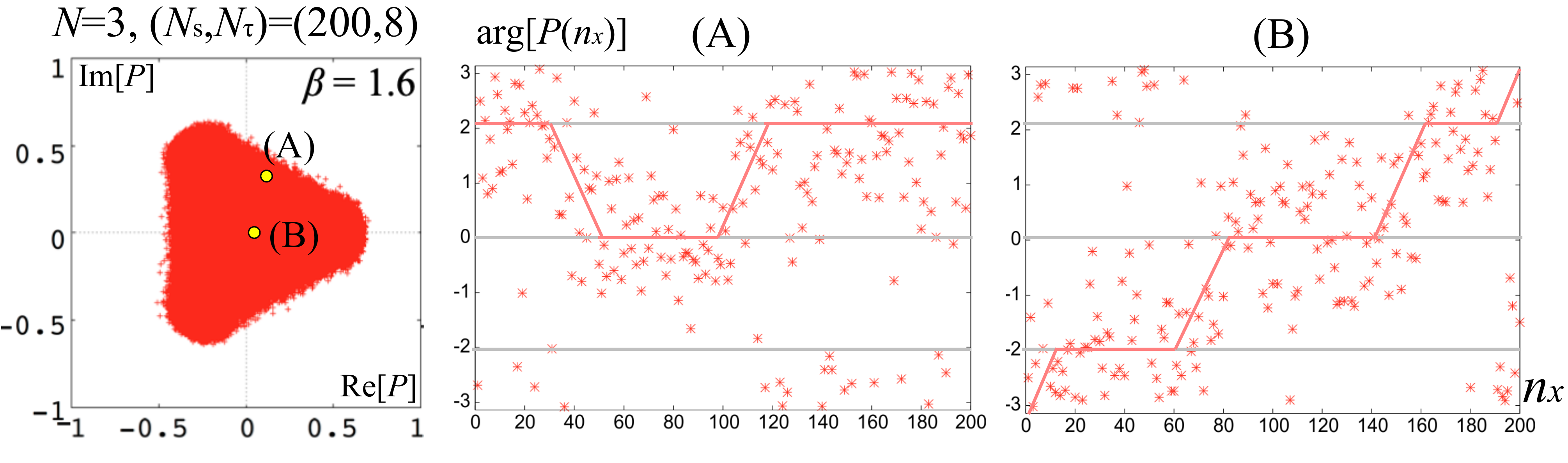}
 \caption{(Left): Distribution plot of the Polyakov loop for $N=3$, $\beta=1.6$ with $(N_{s}, N_{\tau})=(200,8)$ for the $\mathbb Z_{3}$-TBC with the two configurations (A) and (B) pointed.
(Center)(Right): The position dependences of ${\rm arg} [P(n_{x})]$ on $1\leq n_{x} \leq N_{s}$ for the two selected configurations (A) and (B) in the distribution plot. We show the three vacua by gray lines and the estimated vacuum transitions by red lines.
(A) corresponds to a bion while (B) to three fractional instantons.}
\label{fig:frac_Z3}
\end{figure}

\begin{figure}[t]
 \includegraphics[width=1.0\linewidth]{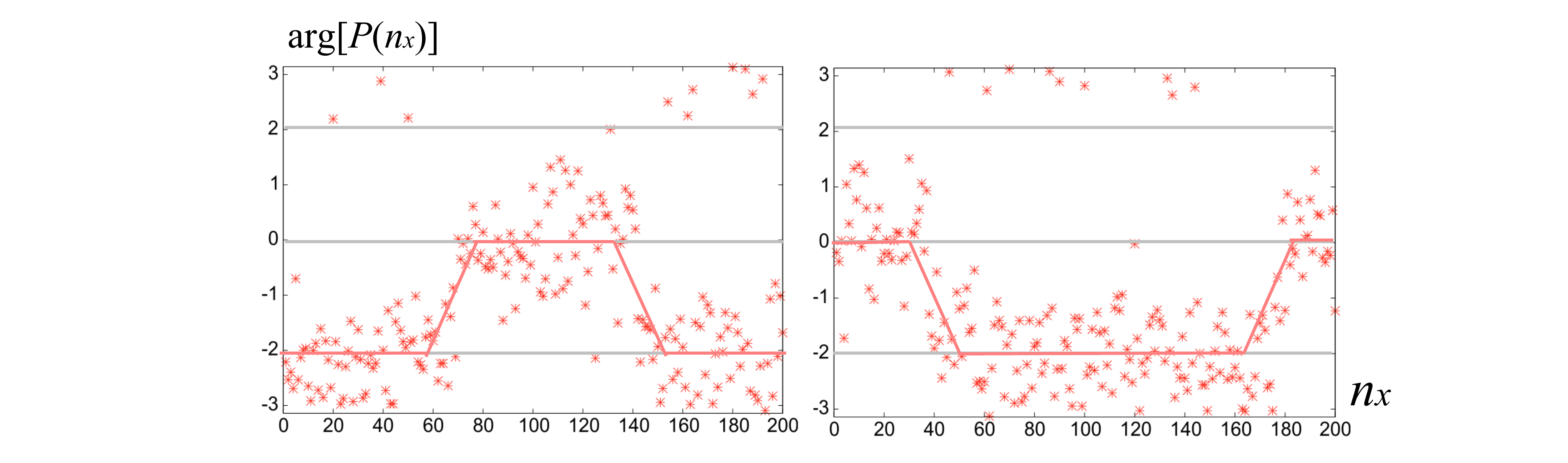}
 \caption{(Left)(Right): The position dependences of ${\rm arg} [P(n_{x})]$ on $1\leq n_{x} \leq N_{s}$ for the two selected configurations (A) and (B) in the distribution plot for $N=3$, $\beta=3.0$ with $(N_{s}, N_{\tau})=(200,8)$ for the $\mathbb Z_{3}$-TBC. We show the three vacua by gray lines and the speculated vacuum transitions by red lines. Both correspond to bion configurations.}
\label{fig:frac_Z3_b30}
\end{figure}


\subsection{Fractional instantons}

We next focus on each of the configurations constituting the $N$-sided polygon-shaped distributions in Fig.~\ref{fig:dist_ZN}.
As we have discussed in Sec.~\ref{sec:CP}, the fractional instantons work to stabilize the $\Z_N$-symmetric vacuum by causing the transition among the classical $N$ vacua.
For given configuration,
whether or not it is a fractional instanton configuration 
can be checked 
by looking at the $n_{x}$ dependence of ${\rm arg} [P(n_{x})]$, which is defined as
\beq
P(n_{x})\equiv \prod_{n_\tau} \lambda_{n,\tau}.
\eeq
$n_{x}$ is the lattice coordinate for $S_{s}^{1}$(large) direction.
We note that ${\rm arg} [P(n_{x})] \approx \oint A_{\tau} d\tau$ describes the vacuum transition process since the topological charge $Q$ for ${\mathbb R}\times S^{1}$ is given by \cite{Eto:2006mz}
\begin{equation}
Q= {1\over{2\pi}} \int \epsilon_{\mu\nu}\partial_{\mu}A_{\nu} = {1\over{2\pi}}
\left[
\oint A_{\tau}(x,\tau) d\tau \right]^{x=\infty}_{x=-\infty} .
\end{equation}
For our lattice setup on $S_{s}^{1}$(large) $\times$ $S_{\tau}^{1}$(small), 
the PBC is also imposed on $n_{x}$ direction, 
thus the total $Q$ should be an integer \cite{Eto:2007aw}.
However, the transition process can be nontrivial, 
and it can include nontrivial configurations such as bions.

We now pick up two of configurations (A) and (B) constituting the polygon-shaped distribution at $\beta=1.6$ for $N=3$ in Fig.~\ref{fig:frac_Z3}(Left). 
In Fig.~\ref{fig:frac_Z3}(Center), we depict ${\rm arg} [P(n_{x})]$ for the configuration (A), which corresponds to one point between the two adjacent ${\mathbb Z}_{3}$ vacua in Fig.~\ref{fig:frac_ZN}(Left). 
In Fig.~\ref{fig:frac_Z3}(Right), we depict ${\rm arg} [P(n_{x})]$ for the configuration (B), which corresponds to one point near the origin in Fig.~\ref{fig:frac_Z3}(Left).
The three classical vacua correspond to ${\rm arg} [P(n_{x})] =0, \pm 2\pi/3$ and we exhibit them by three gray lines in the figures.
The configuration (A) in Fig.~\ref{fig:frac_ZN}(Center) is interpreted to be composed of one fractional instanton and one fractional anti-instanton, so that the total topological charge is zero. It is called a bion configuration.
On the other hand, the configuration (B) in Fig.~\ref{fig:frac_ZN}(Right) is interpreted to be composed of three fractional instantons constituting a single instanton.
From these results, we conclude that the polygon-shaped distribution of the Polyakov loop leading to the ${\mathbb Z}_{N}$-symmetric vacuum is realized by the tunneling transition among $N$ classical vacua due to the fractional instantons.

Since the relatively low-$\beta$ configurations suffer from large fluctuations, 
it is not easy to identify fractional instanton configurations without the auxiliary lines as shown in Fig.~\ref{fig:frac_Z3}(Center)(Right).
For high-$\beta$, these configurations, if exist, become clean because of less fluctuations. 
Although the polygon-shaped distribution tends to be broken at extremely high $\beta$ due to the shortage of statistics as we have discussed, the fractional instanton configuration can be found at certain probability even at quite high $\beta$.
For instance, we find the fractional instanton configurations at $\sim 10\%$ probability at $\beta=3.0$ for $N=3$.
We depict ${\rm arg} [P(n_{x})]$ for two of such configurations at $\beta=3.0$ in Fig.~\ref{fig:frac_Z3_b30}(Left)(Right).
We can identify the fractional instanton configurations quite easily due to small fluctuations for these cases.

\begin{figure}[t]
 \includegraphics[width=0.9\linewidth]{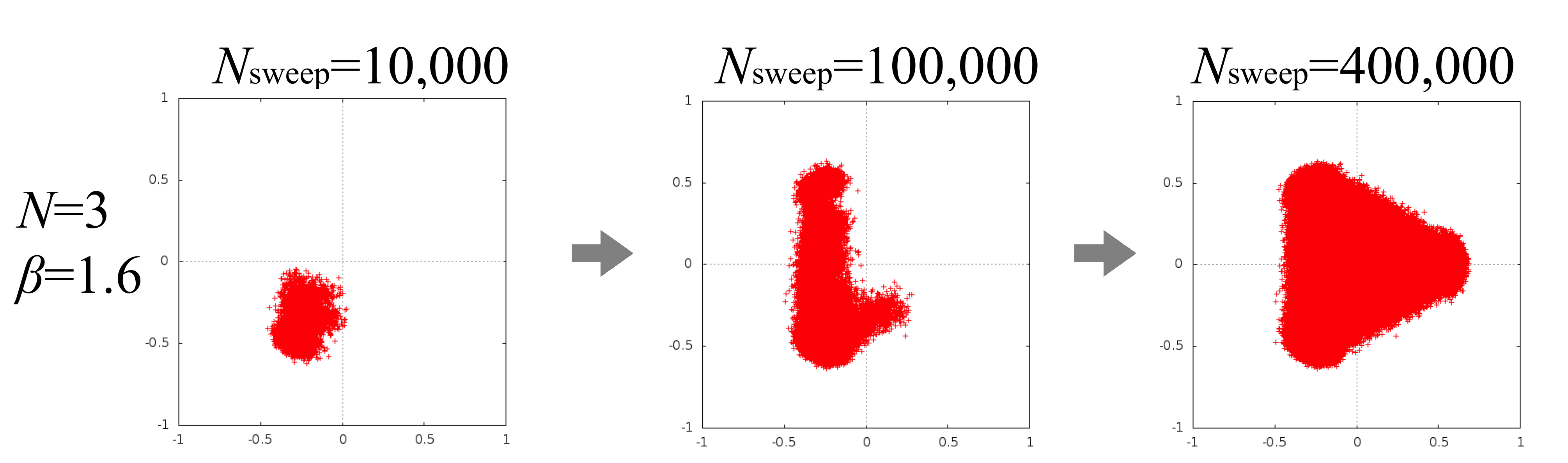}
 \includegraphics[width=0.9\linewidth]{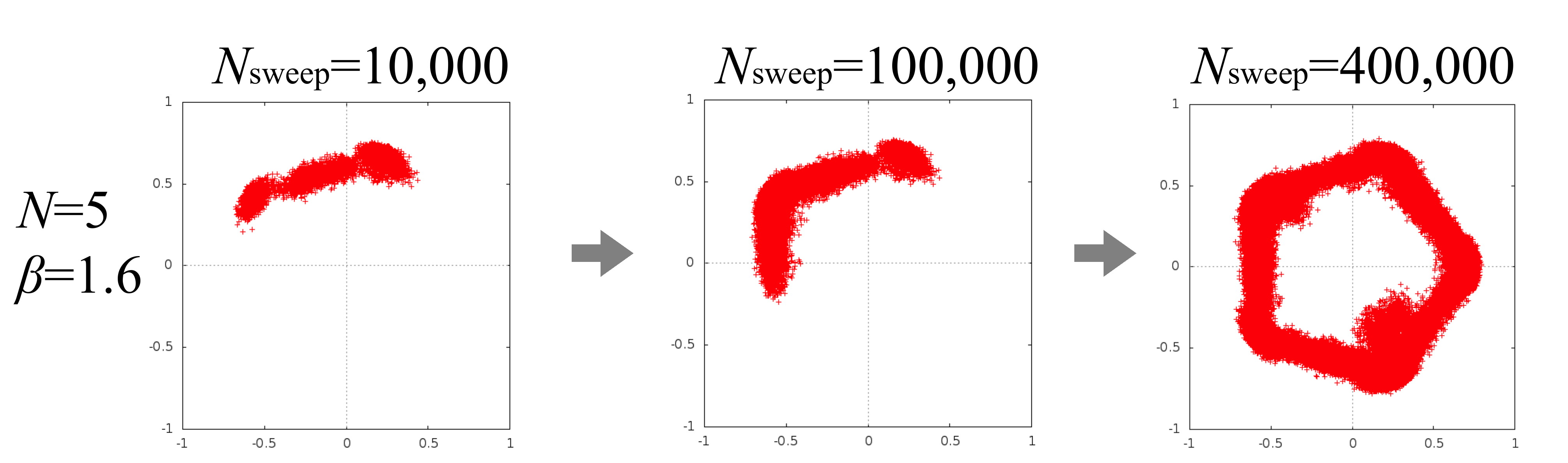}
 \includegraphics[width=0.9\linewidth]{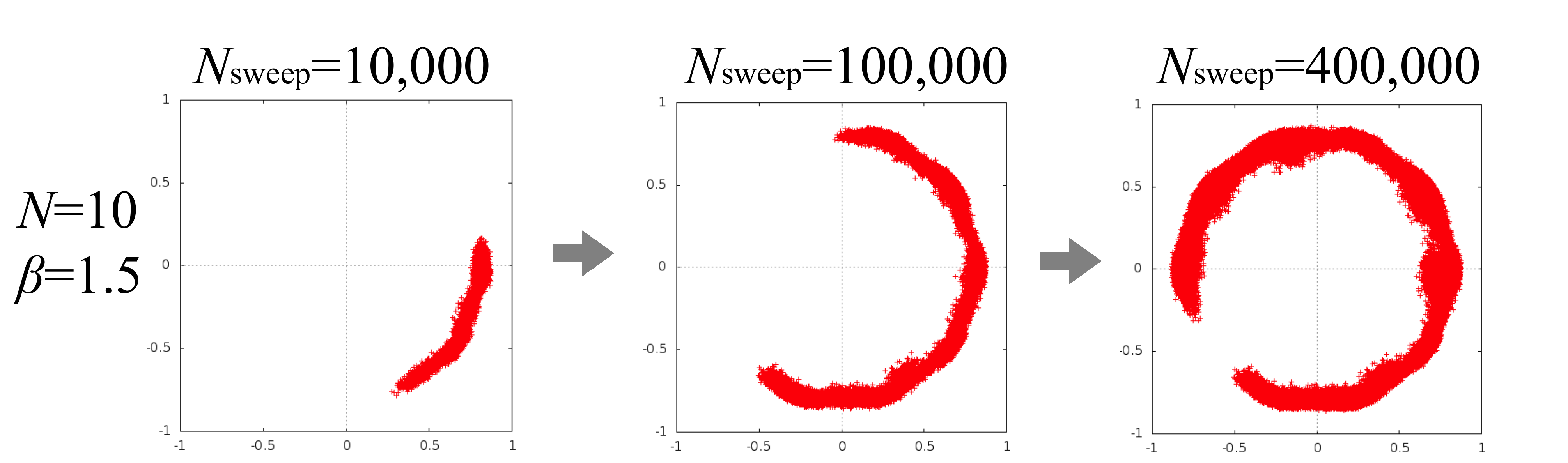}
 \includegraphics[width=0.9\linewidth]{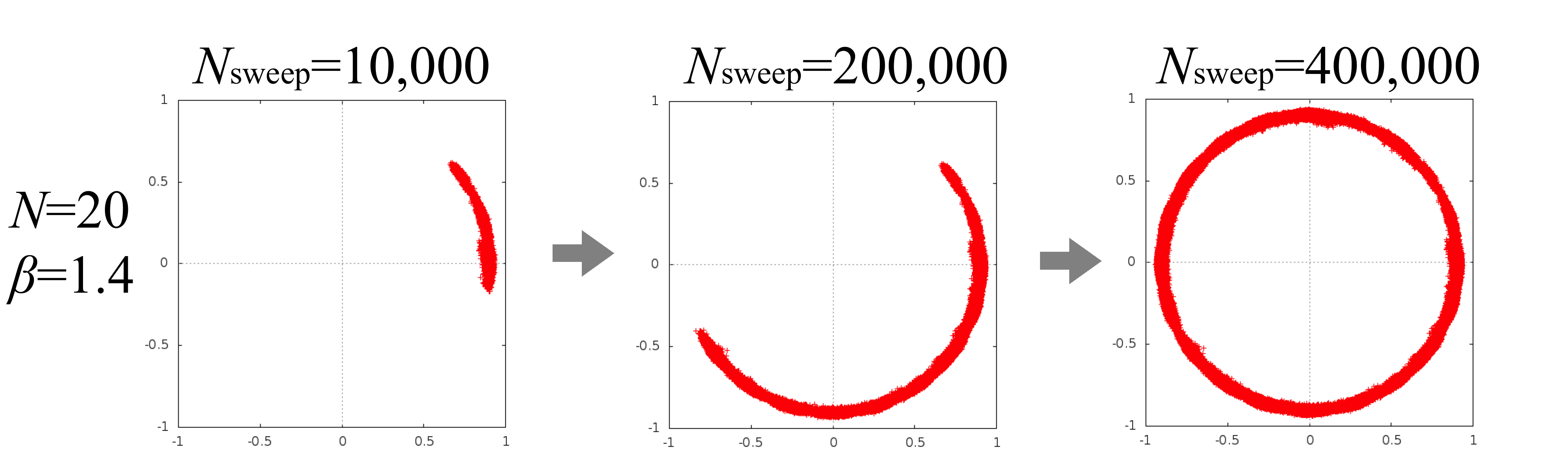}
 \caption{Distribution plots of the Polyakov loop for $N=3,5$ ($\beta=1.6$), $N=10$ ($\beta=1.5$) and $N=20$ ($\beta=1.4$) with $(N_{s},N_{\tau})=(200,8)$ and different numbers of samples in the ${\mathbb Z}_{N}$-TBC. Regular $N$-sided polygon shapes appear as the statistics increases.}
\label{fig:N_stat}
\end{figure}

\begin{figure}[t]
 \includegraphics[width=0.48\linewidth]{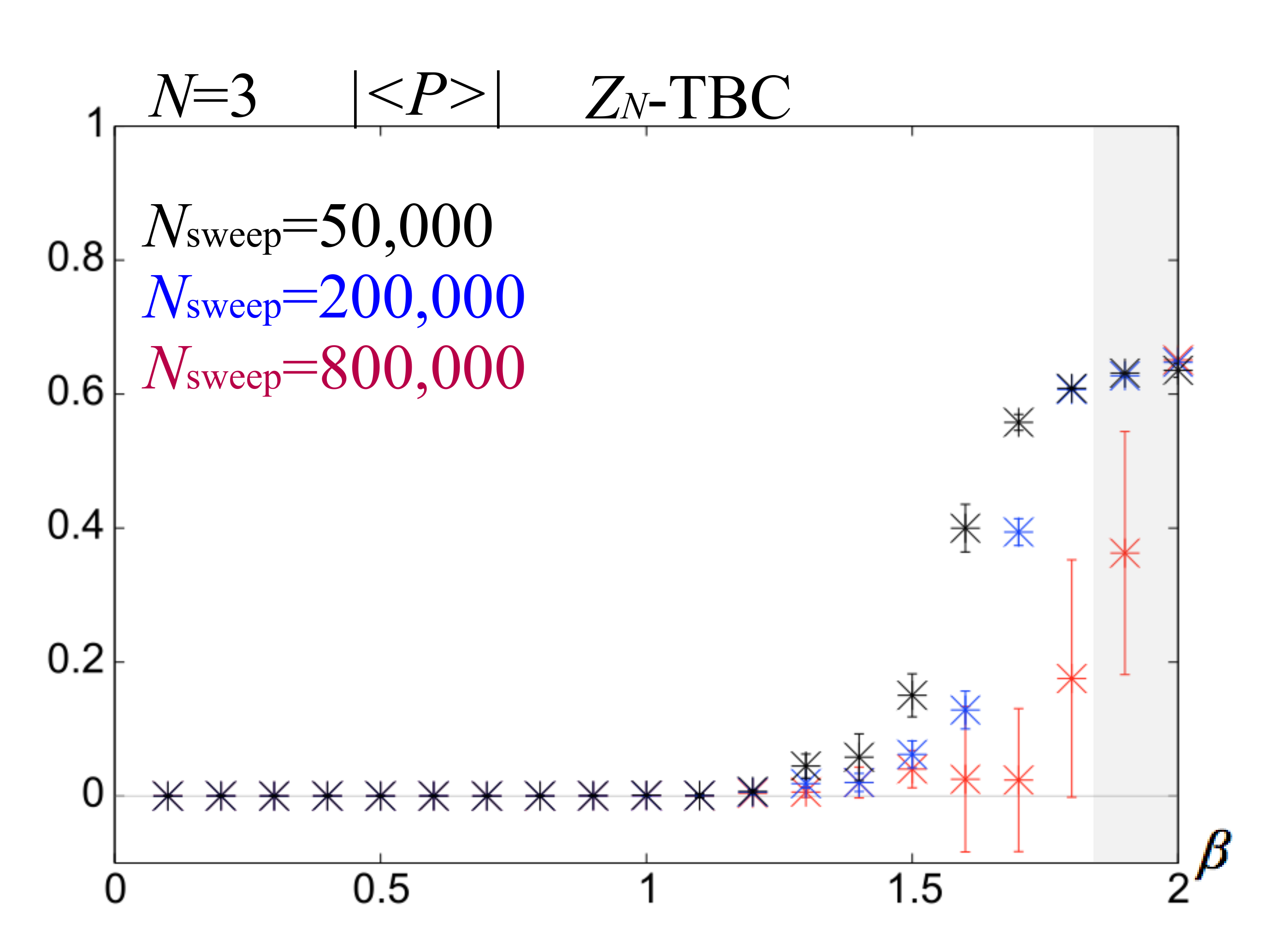}
 \includegraphics[width=0.48\linewidth]{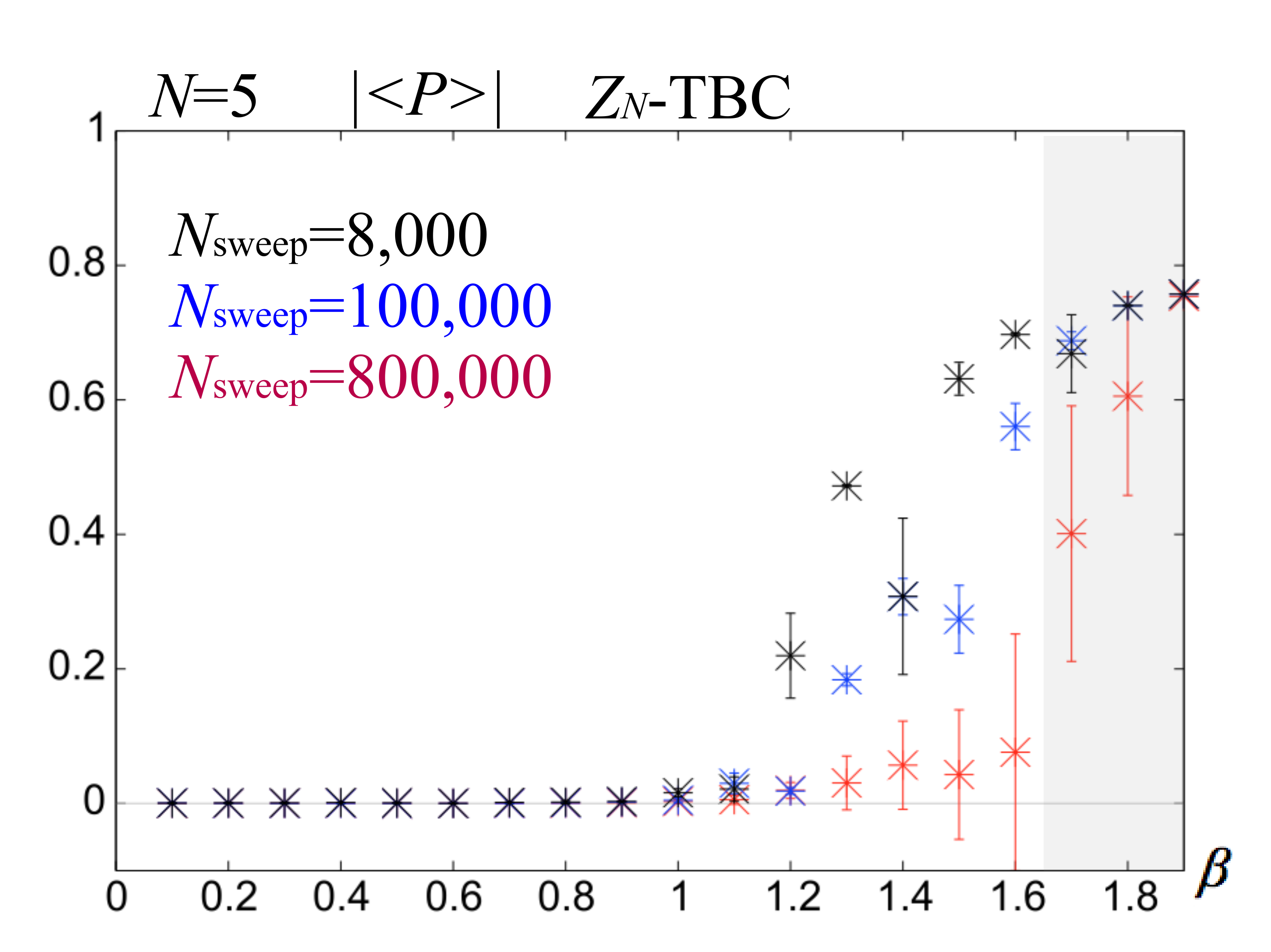}
  \includegraphics[width=0.48\linewidth]{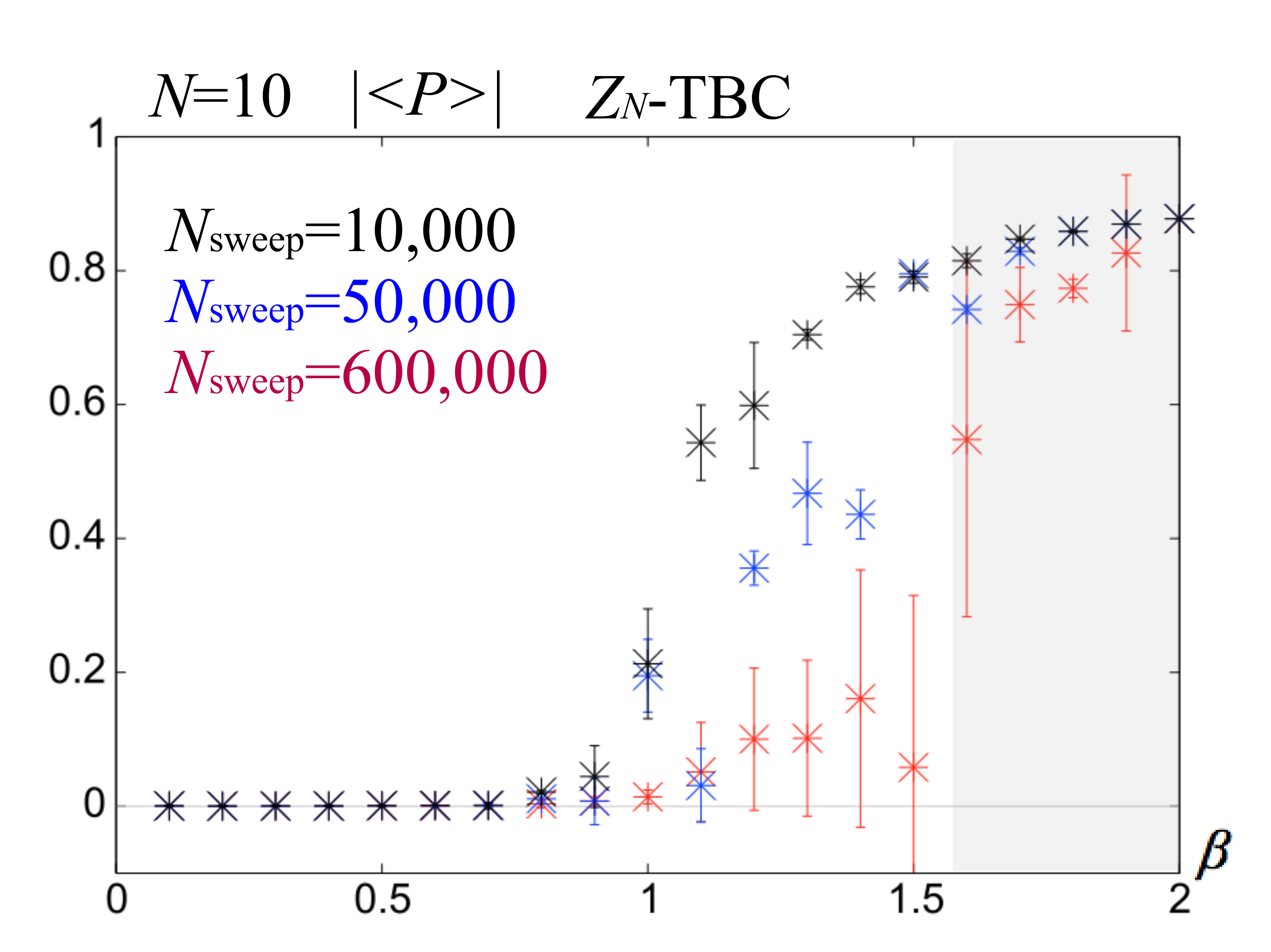}
   \includegraphics[width=0.48\linewidth]{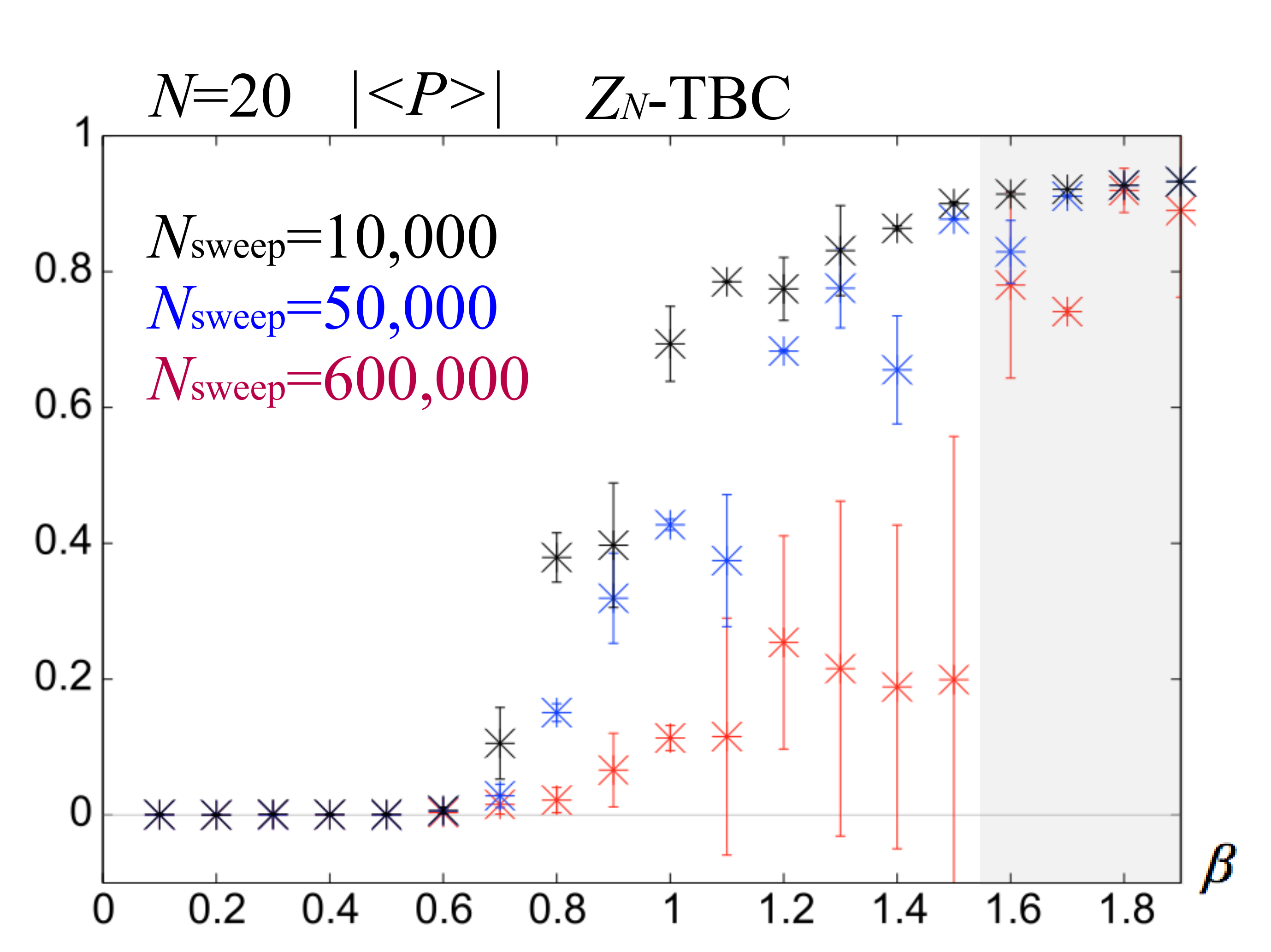}
 \caption{Expectation values of the Polyakov loop $|\langle P \rangle|$ for $N=3,5,10,20$ and $(N_{s},N_{\tau})=(200,8)$ with different numbers of samples in ${\mathbb Z}_{N}$-TBC. The gray-shaded parts indicate the region at which the number of samples are less than the auto-correlation time. 
 In such a high $\beta$ region, larger statistics tend to give smaller expectation values.}
\label{fig:N_statP}
\end{figure}


\subsection{Dependence on the statistics $N_{\rm sweep}$}

We now study how the distribution and expectation values of the Polyakov loop depend on the number of Monte Carlo samples $N_{\rm sweep}$. 

In Fig.~\ref{fig:N_stat}, we show the distribution plots of the Polyakov loop for $N=3,5$ ($\beta=1.6$, $N_{\rm sweep}=10,000,\,\,100,000,\,\,400,000$),  $N=10$ ($\beta=1.5$, $N_{\rm sweep}=10,000,\,\,100,000,\,\,400,000$), and $N=20$ ($\beta=1.4$, $N_{\rm sweep}=10,000, \,\,200,000,\,\,400,000$).
We find that the regular-polygon shape of the Polyakov-loop distribution appears as $N_{\rm sweep}$ increases.

In Fig.~\ref{fig:N_statP}, the absolute values of expectation values of the Polyakov loop $|\langle P\rangle|$ as a function of $\beta$ with three different $N_{\rm sweep}$.
To estimate the Jackknife error, we set the bin-size larger than the auto-correlation time for the case with the largest number of samples while we adopt $N_{\rm bin}=1000$ for the other cases for simplicity.
By increasing $N_{\rm sweep}$, the values in high $\beta$ regions rapidly decrease and get close to zero. 
Furthermore, even the values in much higher $\beta$ regions (gray-shaded regions in the figure), in which the statistics are expected to be less than the auto-correlation times, decrease as $N_{\rm sweep}$ increase.
From this observation, we speculate that larger statistics lead to $|\langle P\rangle|\sim 0$ even in this $\beta$ region.
More generically speaking, the $\Z_N$-symmetric vacuum, where $|\langle P\rangle| $ is consistent with zero, could be observed even at extremely high $\beta$ if we generate large enough number of samples beyond the auto-correlation time.

We now make comments on the reason why the auto-correlation time increases and why we need larger statistics for higher $\beta$. 
The theoretical reason is a rapid decrease of the transition probability among the $N$ classical vacua.
The transition probability between adjacent $N$ classical vacua due to $1/N$ fractional instantons with the action $S={2\pi\over{Ng^{2}}}$ is expressed as
\begin{equation}
{\rm Transition \,\,probability} \,\propto\,\left[\exp\left( -{2\pi\over{Ng^{2}}} \right)\right]^{2} \,.
\end{equation}
Although $g^{2}$ should be the renormalized coupling, we can approximate it by the bare coupling $g_{0}^{2}={1\over{N\beta}}$ in the weak-coupling limit (corresponding to small $L_{\tau}$ or large $\beta$ limit). 
Therefore, the transition probability exponentially decreases when $\beta$ increases.
In the Monte Carlo simulation, it means that we need exponentially large $N_{\rm sweep}$ to observe a true quantum vacuum, which is expected to have a small expectation value of the Polyakov loop ($|\langle P\rangle|\sim0$). This is one of possible reasons why the auto-correlation time surges for larger $\beta$.
It is also notable that the vacuum ``freezes" for very high $\beta$ and the auto-correlation time seems to get small in appearance because of the small transition rate among the classical vacua.
However it is as an artifact and a large auto-correlation time should be observed if one adopts exponentially large statistics.
In other words, the statistical errors of $|\langle P\rangle|$ in extremely high $\beta$ regions may be underestimated in the simulation due to the freezing of the vacuum.

Since we do not have enough statistics at $\beta>2.0$ for any $N$ in our simulations,
we cannot have a strong conclusion on the adiabatic continuity of the $\Z_{N}$-symmetry for the whole $\beta$ regime. However, we emphasize that the $\Z_{N}$-symmetric regions exist even in $\beta_c <\beta$, where the IR scale is no longer an ordinary confinement scale as shown in Sec.~\ref{sec:Ploop}.
Furthermore, the results in this subsection imply that we may be able to observe the adiabatic continuity by adopting an exponentially large number of samples $N_{\rm sweep}$. 
This speculation will be also supported by the arguments in 
Sec.~\ref{sec:large-volume}.

\begin{figure}[t]
\includegraphics[width=0.99\linewidth]{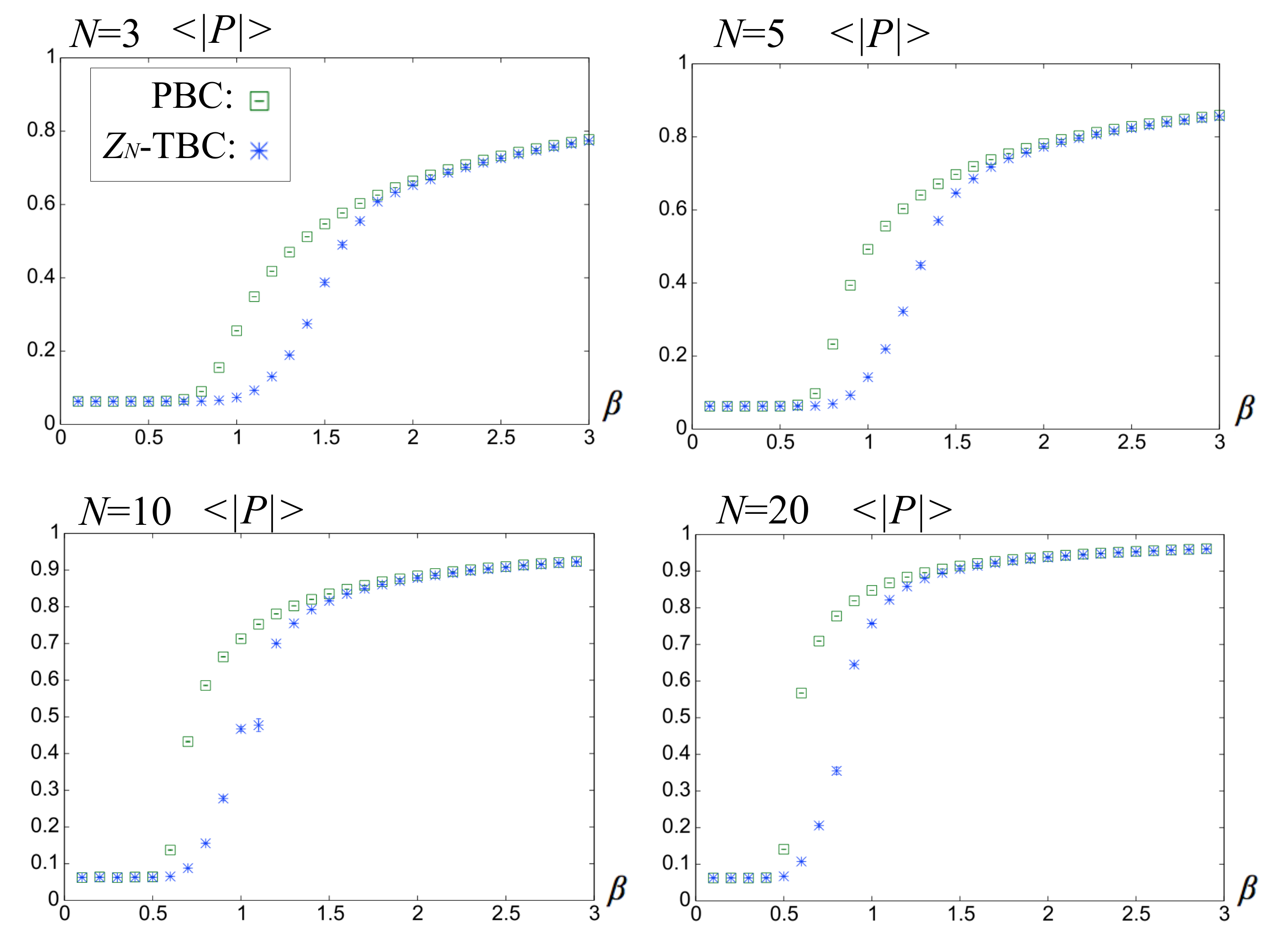}
 \caption{
The expectation values of the absolute values of the Polyakov loop for the ${\mathbb Z}_{N}$-TBC and PBC with $N=3,5,10,20$.}
\label{fig:Ploopab_ZNvsPBC}
\end{figure}


\subsection{Comparison between $\Z_N$-twisted boundary condition and periodic boundary condition}

We next make comparison between the expectation values of the absolute values of the Polyakov loop $\langle |P| \rangle$  for the $\Z_N$-TBC and PBC for $N=3,5,10,20$.
We again note that this quantity $\langle |P| \rangle$ is not the order parameter of the $\Z_N$ symmetry, but the indicator of the expanse of the Polyakov-loop distribution.
The reason why we here discuss $\langle |P| \rangle$ instead of $|\langle P \rangle |$ is to show the absence of phase transitions (or the absence of sudden expanses of the distribution) in the $\Z_N$-twisted model.

In the previous work \cite{Fujimori:2019skd}, it was shown that the $\beta$ dependence of $\langle |P| \rangle$ undergoes a crossover behavior in the $\C P^{N-1}$ model with PBC.
We here make use of this fact and make comparison between $\langle |P| \rangle$ of the $\Z_N$-TBC and PBC.
If we find gentler increase of $\langle |P| \rangle$ in the $\Z_N$-TBC than that in the PBC as $\beta$ increases, we can conclude the absence of phase transitions between large-$L_{\tau}$ and small-$L_{\tau}$ regions at least for this quantity.

We depict $\langle |P| \rangle $ for the $\mathbb Z_{N}$-TBC and PBC in Fig.~\ref{fig:Ploopab_ZNvsPBC}. 
The reason why we plot them from small to quite large $\beta$ ($\beta <3.0$) is that the auto-correlation time for this quantity is much smaller than that for $|\langle P \rangle |$ and the results of the simulations are reliable.
It is obvious that, as $\beta$ increases, $\langle |P| \rangle $ for the $\Z_N$-TBC gets larger more gradually than that for the PBC.
It means that $\langle |P| \rangle $ for the $\Z_N$-TBC undergoes a crossover behavior between large-$L_{\tau}$ and small-$L_{\tau}$ regions, thus there are no phase transitions regarding $\langle |P| \rangle $ in this model.
We consider that this is an affirmative fact for the adiabatic continuity of the vacuum since the crossover behavior of $\langle |P| \rangle $ is consistent with the absence of sudden change of vacuum structure.


\subsection{Large-volume simulation}\label{sec:large-volume}
 
In this subsection, we show results of the simulation with larger volume $(N_{s}, N_{\tau})=(400,12)$ for $N=3$.
In particular, we perform an exploratory simulation at very high $\beta$ as $\beta=4.0$ with $N_{\rm sweep} =198,000$ and obtain distribution plots of the Polyakov loop. 
Since an auto-correlation time for $|\langle P \rangle|$ at $\beta=4.0$ is speculated to be quite large, 
this $N_{\rm sweep}$ is probably insufficient.
Therefore the reliability of results in this simulation is lower than those in other parts of this paper,
and we do not show a result of $|\langle P \rangle|$ but just discuss one of distribution plots.
This is why we refer to the simulation in this subsection as an ``exploratory" one.

\begin{figure}[t]
\includegraphics[width=0.8\linewidth]{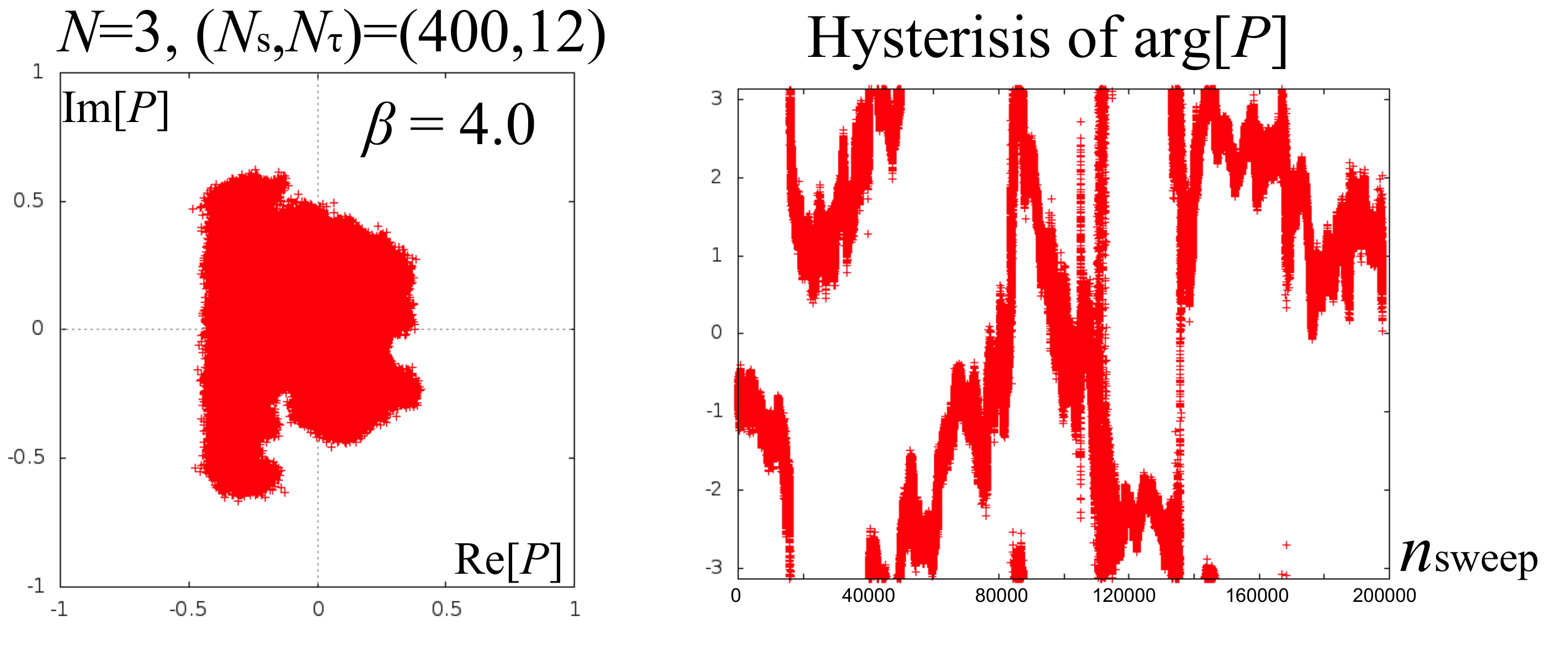}
\caption{(Left): Distribution plot of the Polyakov loop for $N=3$, $\beta=4.0$ with $(N_{s}, N_{\tau})=(400,12)$ for $\mathbb Z_{N}$-TBC. (Right): The corresponding hysterysis of ${\rm arg}[P]$ for $N=3$, $\beta=4.0$ with $(N_{s}, N_{\tau})=(400,12)$. }
\label{fig:Lvolume_ZN}
\end{figure}

In Fig.~\ref{fig:Lvolume_ZN}(Left), we depict one of the distributions plots of the Polyakov loop, which has a partially broken regular-triangle shape.
We check that the history of ${\rm arg} [P]$ for this distribution is quite random in Fig.~\ref{fig:Lvolume_ZN}(Right), implying that the transition among the ${\mathbb Z}_{N}$ vacua occurs perpetually.
This result implies that, even at quite high $\beta$, the regular-polygon distribution of the Polyakov loop appears with a certain probability ($\sim 5 \%$) at least for a larger volume $(N_{s}, N_{\tau})=(400,12)$.

\begin{figure}[t]
 \includegraphics[width=1.0\linewidth]{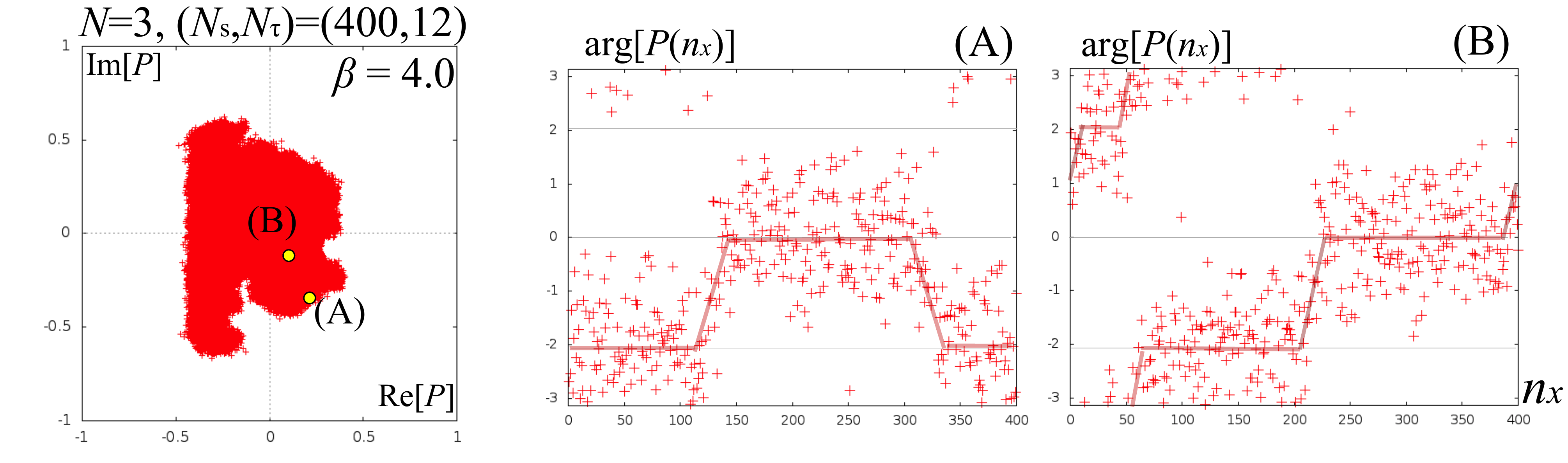}
 \caption{(Left): Distribution plot of the Polyakov loop for $N=3$, $\beta=4.0$ with $(N_{s}, N_{\tau})=(400,12)$ for $\mathbb Z_{N}$-TBC with the two configurations (A)(B) pointed.
(Center)(Right): Position dependences of ${\rm arg} [P(n_{x})]$ on $1\leq n_{x} \leq N_{s}$ for the two selected configurations (A)(B) in the distribution plot. We show the three vacua by gray lines and the speculated vacuum transitions by red lines.
(A) corresponds to a bion while (B) to three fractional instantons.}
\label{fig:frac_ZN}
\end{figure}

In this simulation, the fractional instanton configurations are also observed. 
We pick up two configurations (A) and (B) corresponding to the two points shown in Fig.~\ref{fig:frac_ZN}(Left).
In Fig.~\ref{fig:frac_ZN}(Center), ${\rm arg} [P(n_{x})]$ for the configuration (A) corresponding to one point between the two adjacent ${\mathbb Z}_{3}$ vacua in Fig.~\ref{fig:frac_ZN}(Left) is depicted. 
In Fig.~\ref{fig:frac_ZN}(Right), ${\rm arg} [P(n_{x})]$ for the configuration (B) corresponding to one point near the origin in Fig.~\ref{fig:frac_ZN}(Left) is depicted.
The configuration (A) in Fig.~\ref{fig:frac_ZN}(Center) is composed of one fractional instanton and one fractional anti-instanton constituting a bion, while the configuration (B) in Fig.~\ref{fig:frac_ZN}(Center) is composed of three fractional instantons constituting a single instanton.
These results implies that the quantum transition can be caused by fractional instantons even at quite high $\beta$.

\section{Pseudo-entropy density}
\label{sec:entropy}

We now study the ``pseudo-entropy", 
which is a counterpart of 
the thermal entropy in the PBC case and
is related to the degrees of freedom of the system.
It is defined in the same way as the thermal entropy
except that the $\Z_N$-TBC is imposed instead of the PBC
along the Euclidean time direction. 
Although it has different properties 
such as non-positive definiteness, 
we call it ``pseudo-entropy density" for $\Z_N$-TBC, 
since it still carries similar properties 
to the thermal entropy. 
In terms of the energy-momentum tensor (EMT) $T_{xx},\,T_{\tau\tau}$, 
the pseudo-entropy density $s$ in the large volume limit 
is given by
\begin{equation}
s =  \langle T_{xx}-T_{\tau\tau}\rangle /T, \hspace{10mm}
\mbox{with~~$T\equiv 1/L_{\tau}$},
\end{equation}
where the $\Z_N$-TBC along the $\tau$ direction is imposed. 
On the lattice, EMT is defined \cite{Asakawa:2013laa} as
\beq
&&T_{xx}\,=\,2 N\beta (2- \bar{\phi}_{n+x}\cdot \phi_{n}\lambda_{n,x} 
- \bar{\phi}_{n}\cdot \phi_{n+x}\bar{\lambda}_{n,x}) -(\mbox{trace part}),
\\
&&T_{\tau\tau}\,=\,2 N\beta (2- \bar{\phi}_{n+\tau}\cdot \phi_{n}\lambda_{n,\tau} 
- \bar{\phi}_{n}\cdot \phi_{n+\tau}\bar{\lambda}_{n,\tau}) -(\mbox{trace part}),
\eeq
where the vacuum expectation value of the trace part is subtracted \cite{Makino:2014sta,Makino:2014cxa}.
We will adopt the bare coupling constant to calculate these quantities since it well approximates the renormalized coupling in the weak coupling regime. 

\begin{figure}[t]
 \includegraphics[width=0.49\linewidth]{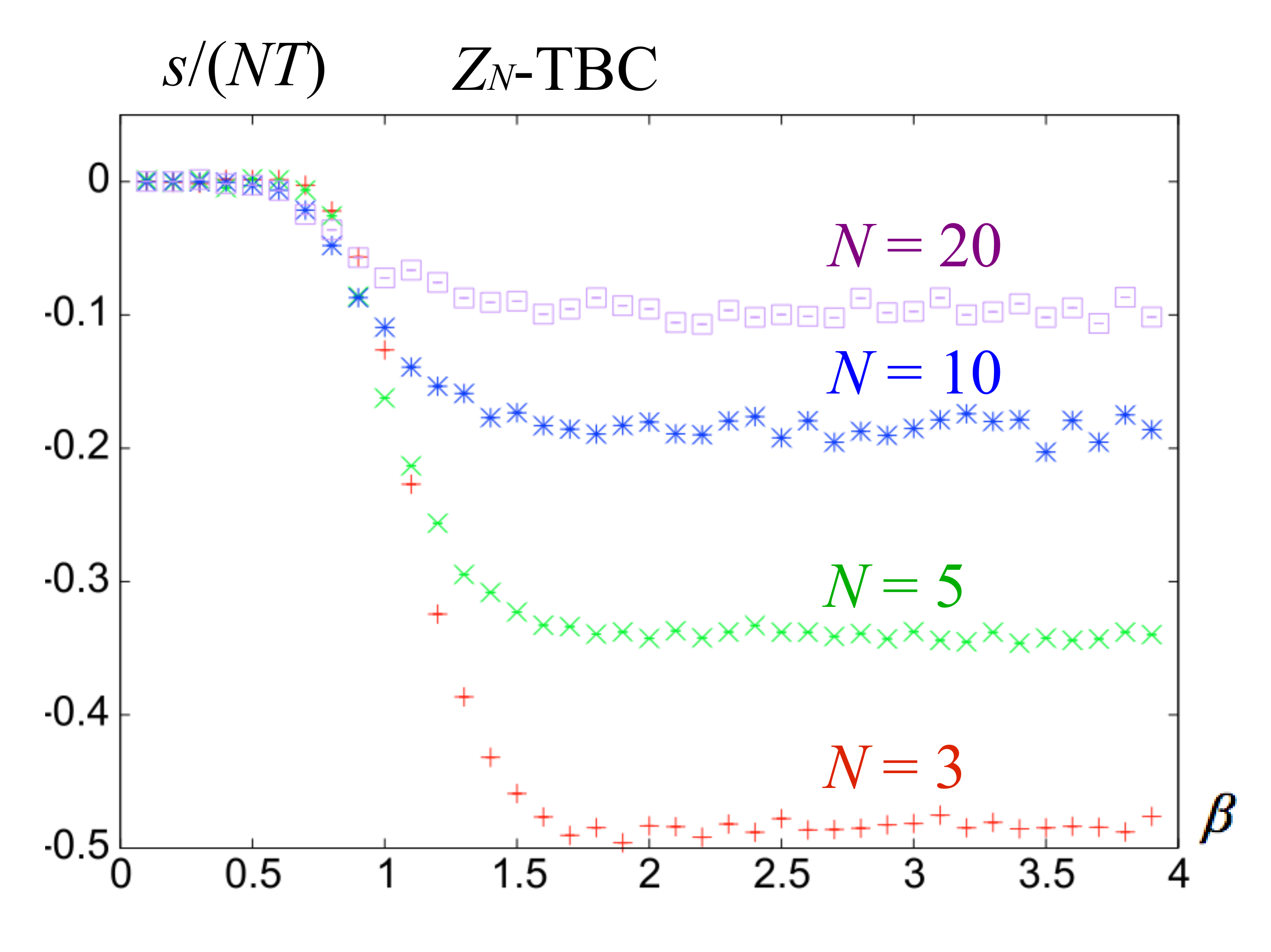}\,\,\,
 \includegraphics[width=0.49\linewidth]{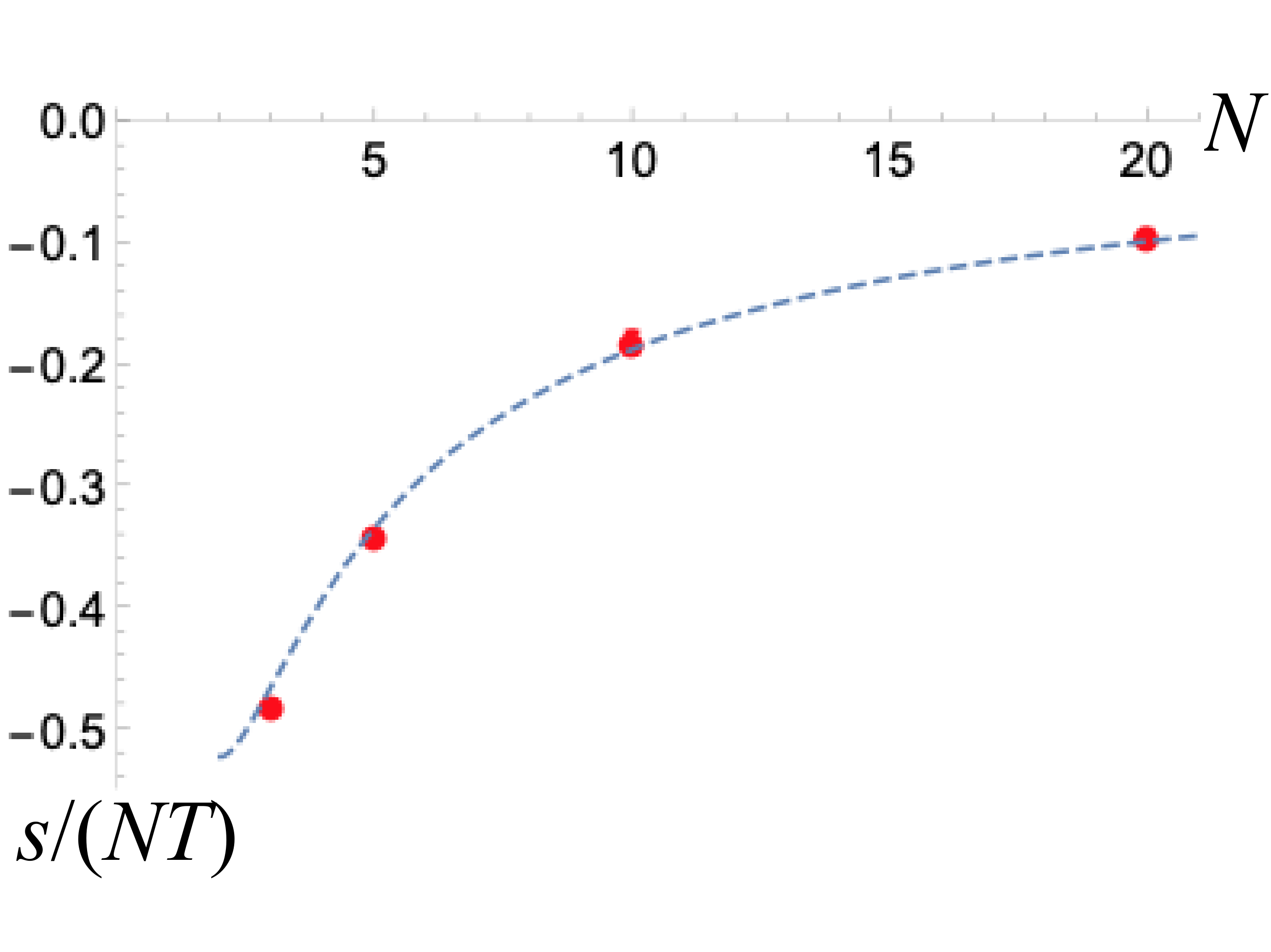} 
 \caption{(Left): Pseudo-entropy density ($s/(NT)= N_{\tau}^{2}\langle T_{xx}-T_{\tau\tau}\rangle/N$) for $N=3,5,10,20$. (Right): The averaged values of pseudo entropy density for $N=3,5,10,20$ ($\beta=3.0-3.9$) after they are well saturated. The broken curve stands for the analytically speculated value $-{2\pi\over{3N^{2}}}(N-1)$ up to ${\mathcal O}(1/N^2)$ corrections.}
\label{fig:entropy}
\end{figure}

Let us discuss the simulation results of the pseudo-entropy 
density $s/(NT)$.
The $\beta$ dependences of the pseudo-entropy density for $N=3,5,10,20$ are depicted in Fig.~\ref{fig:entropy}(left).
The pseudo-entropy density becomes non-zero around a certain $\beta$ and monotonically decreases.
In the high-$\beta$ regime, 
the $\beta$ dependence gets gentler, 
then the value reaches a plateau for each $N$.
We average the saturated values between $3.0 \le \beta \le 3.9$ and obtain 
$s_{N=3}/(3T)=-0.482(4)$, $s_{N=5}/(5T)=-0.342(3)$, $s_{N=10}/(10T) =-0.183(9)$, 
$s_{N=20}/(20T)= -0.096(6)$.
We depict them in Fig.~\ref{fig:entropy}(right)
and find that the values are consistent with 
the pseudo-entropy density for $N-1$ free massive scalar fields
(see App.\,\ref{sec:free_energy})
\beq
s ~=~-{2\pi\over{3L_{\tau} N}} (N-1)\,+\,{\mathcal O}(m)\,.
\eeq
which is depicted as the broken curve.
For small $N$ such as $N=3,5$ in the figure, 
small deviations from the analytical result are found.
We speculate that it indicates $1/N^2$ corrections to the leading-order results.

It is worth noting that, in the large-N limit, 
the pseudo-entropy in the high $\beta$ regime 
seems to go to zero.
Thus, the difference between $T_{xx}$ and $T_{\tau \tau}$ disappears even though the anisotropy between $x$ and $\tau$ directions is large. It suggests the volume independence of the EMT and the action density.
This speculated property in the large-$N$ limit is consistent with the analytical result proposed in Ref.~\cite{Sulejmanpasic:2016llc}.

It is also notable that the pseudo-entropy densities do not exhibit any special behaviors such as phase transition at high $\beta$.
We consider that these results of the pseudo-entropy density also support the adiabatic continuity of the model between $\R \times S^{1}$ and $\R^{2}$ although this quantity is not sensitive to the breaking of the $\Z_N$ symmetry.

In the end of this section, we make a brief comment on the pseudo-entropy in the QCD-like models.
The pseudo-entropy can be also calculated in the lattice simulation of $N$-flavor QCD with the $\Z_N$ twisted boundary condition ($\Z_N$-QCD) \cite{Iritani:2015ara,Misumi:2015hfa}. 
The question whether it is negative or not in $\Z_N$-QCD is quite nontrivial and will be addressed in future works.


\section{Summary and discussion}
\label{sec:summary}

In this paper, we have reported the Monte Carlo simulations for the ${\mathbb C}P^{N-1}$ model on $S_{s}^{1}({\rm large}) \times S_{\tau}^{1}({\rm small})$ with the ${\mathbb Z}_{N}$-TBC, where sufficiently large ratio of the circumferences is taken to approximate the model on $\mathbb R \times S^1$.
We have found that the $\beta$ dependence of expectation values 
of the Polyakov loop 
differs significantly from those for the PBC.
Our results so far seem to
give evidences in support for 
the conjecture of the adiabatic continuity of the vacuum structure.
We here summarize our main results:

\begin{enumerate}

\item
By studying $\langle|P|\rangle$, we find that the $\Z_N$-twisted model has the characteristic $\beta$ denoted as $\beta_{c}$, at which the IR scale of the system changes from $\Lambda_{\C P^{N-1}}$ to $1/L_{\tau}$.

\item
The order parameter of the $\Z_N$ symmetry, $|\langle P\rangle|\sim 0$, continues to be consistent with zero in the both $\beta<\beta_c$ and $\beta > \beta_c$ regions.
It means that the $Z_N$ symmetry is unbroken even in the high $\beta$ regime.
The distribution of the Polyakov loop in $\beta > \beta_c$ forms regular $N$-sided-polygon shapes.

\item
By investigating the dependence of the Polyakov-loop phase on the coordinate of the $S_{s}^{1}$ direction,
the existence of fractional instantons and bions causing the transition among the classical $N$ vacua is verified. We argue that fractional instantons work to stabilize the $\Z_N$ symmetry.

\item
The result on $|\langle P\rangle|$ is unreliable for quite high $\beta$ since the statistics we adopt in the simulation are less than the auto-correlation time.
Even for such a high $\beta$, a regular $N$-sided-polygon shape of the Polyakov-loop distribution tends to be restored and $|\langle P\rangle|$ gets smaller by increasing the 
number of samples.

\item
In a larger-volume simulation, the $N$-sided polygon-shape in the Polyakov loop distribution appears even for extremely high $\beta$ with a certain probability. We note that it is an exploratory calculation and the number of samples is not sufficient.

\item
The $\beta$ dependence of the pseudo-entropy density $s=\langle T_{xx}-T_{\tau\tau}\rangle/T$ implies the absence of phase transitions between the large and small circumference regions. 
Furthermore, the pseudo-entropy density vanishes in the large-$N$ limit.
It is consistent with the volume independence in the whole $\beta$ regime in the large-$N$ limit.

\end{enumerate}

This work helps deepen the understanding on the symmetry and phase diagram of the $\C P^{N-1}$ model on the compactified spacetime with the $\Z_N$-TBC. 
Our main focus was the adiabatic continuity of the ${\mathbb Z}_{N}$ symmetric phase, which is essential to apply the resurgence theory to the model on ${\mathbb R}\times S^{1}$ in the decompactified limit.
Although our results are not conclusive to the conjecture of the adiabatic continuity in the model,
there are various implications for future studies on the topic.

In the end of this paper,
let us introduce our next plan to judge the adiabatic continuity of the $\Z_N$ symmetry in the model:

(1) At quite high $\beta$ (small $L_{\tau}$), we generate configurations which give polygon-shape distribution and a small expectation value of Polyakov loop $|\langle P \rangle| \sim 0$ (e.g. the case of Fig.~\ref{fig:Lvolume_ZN}).

(2) We pick up one of the above configurations and use it as an initial configuration to generate configurations at slightly lower $\beta$.

(3) We repeat this procedure and generate configurations from high to low $\beta$.

(4) We investigate whether $|\langle P \rangle| \sim 0$ continues form high to low $\beta$ (from small to large $L_{\tau}$).

The point is that the generation of configurations in this proposal starts from high $\beta$ while we started it from low $\beta$ in the simulation of the present work.
We can judge the validity of the adiabatic continuity through 
this procedure in principle, although we need to realize the 
``adiabatic" decrease of $\beta$ which is quite hard to achieve 
in lattice simulations.


\begin{acknowledgements}
This work is supported by the Ministry of Education, Culture, 
Sports, Science, and Technology(MEXT)-Supported Program for the 
Strategic Research Foundation at Private Universities ``Topological 
Science" (Grant No. S1511006) 
and 
by the Japan Society for the Promotion of Science (JSPS) 
Grant-in-Aid for Scientific Research (KAKENHI) Grant Number 
(18H01217).
This work is also supported in part by JSPS KAKENHI Grant Numbers 
19K03875 (E.\ I.), 18K03627 (T.\ F.), 19K03817 (T.\ M.), and  16H03984 (M.\ N.).
The work of E.\ I. is supported by the HPCI-JHPCN System Research Project (Project
ID: jh200031).
The work of M.\ N. is also supported in part 
by a Grant-in-Aid for Scientific Research on Innovative Areas 
``Topological Materials Science" (KAKENHI Grant No. 15H05855) 
from MEXT of Japan. 
Numerical simulations were performed on SX-ACE at the Research Center for Nuclear Physics (RCNP), Osaka University and TSC at Hiyoshi department of Physics, Keio University.
\end{acknowledgements}


\appendix

\section{Bin-size dependence of the errors of the Polyakov loop}
\label{sec:bin-size}

In this appendix we show the bin-size $N_{\rm bin}$ dependence of Jacknife errors of the absolute values of the Polyakov-loop expectation values $|\langle P\rangle|$ for $(N_{s},N_{\tau})=(200,8)$ in ${\mathbb Z}_{N}$-TBC. 

\begin{figure}[t]
 \includegraphics[width=0.95\linewidth]{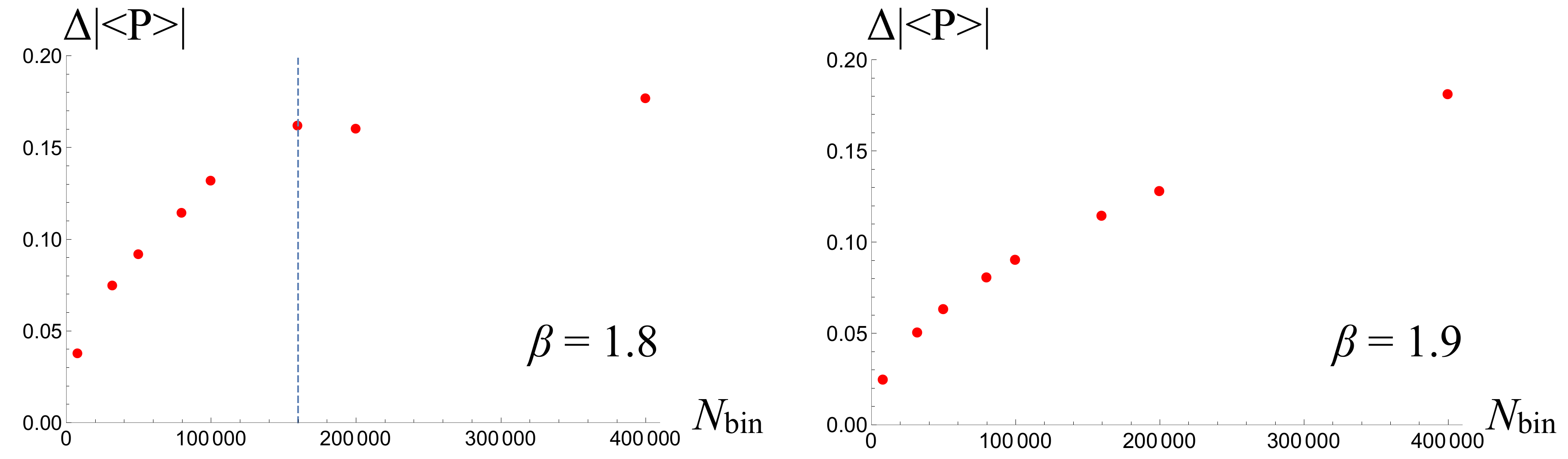}
 \caption{Bin-size dependence of errors of absolute values of the Polyakov-loop expectation values for $N=3$ and $\beta=1.8,1.9$ with $(N_{s},N_{\tau})=(200,8)$ in ${\mathbb Z}_{N}$-TBC.
 The horizontal axis is $N_{\rm bin}$ and the vertical axis is an error of $|\langle P \rangle|$.}
\label{fig:N3Nbin}
\end{figure}

In Fig.~\ref{fig:N3Nbin}, we depict it for $N=3$ and $\beta=1.8,1.9$.
For $\beta=1.8$, the plateau appears around $N_{\rm bin}=160000$.
For $\beta=1.9$, the plateau does not appear within the number of samples we adopt in this work.
Since the value of $N_{\rm bin}$ at which the plateau appears corresponds to auto-correlation time,
we roughly estimate that the simulations for $\beta\leq 1.8$ are performed with sufficient statistics beyond the auto-correlation times.
For higher $\beta$, we do not reach a plateau within the statistics, thus we consider that they are less than the auto-correlation time and the simulation results are not reliable.

\begin{figure}[t]
 \includegraphics[width=0.95\linewidth]{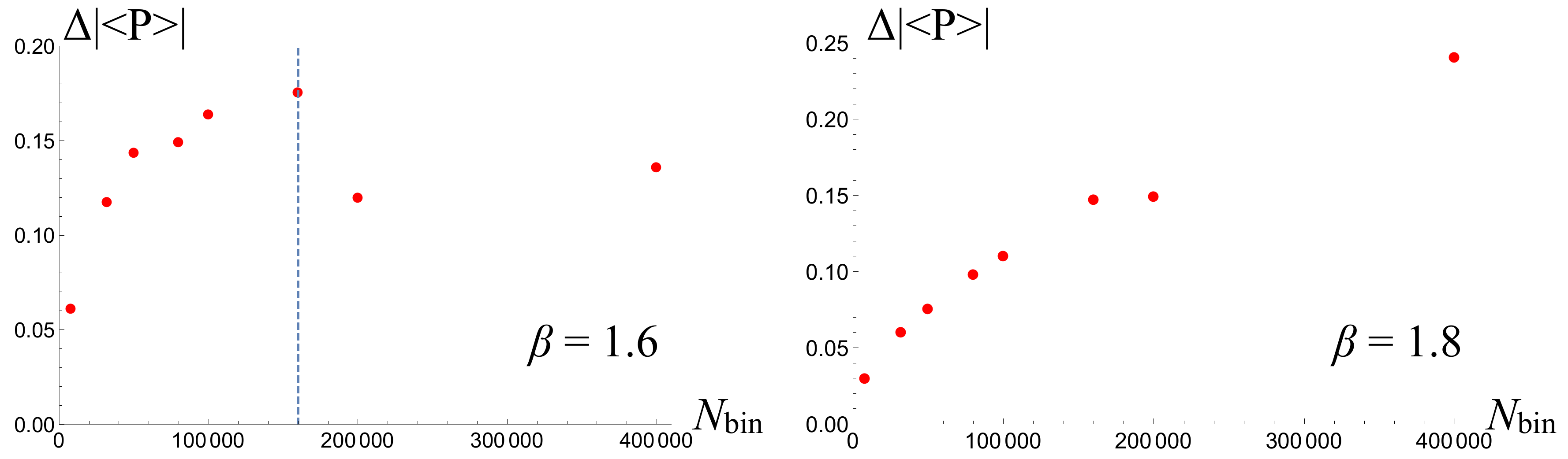}
 \caption{Bin-size dependence of errors of absolute values of the Polyakov-loop expectation values for $N=5$ and $\beta=1.6,1.8$ with $(N_{s},N_{\tau})=(200,8)$ in ${\mathbb Z}_{N}$-TBC.
 The horizontal axis is $N_{\rm bin}$ and the vertical axis is an error of $|\langle P \rangle|$.}
\label{fig:N5Nbin}
\end{figure}

In Fig.~\ref{fig:N5Nbin}, we depict the bin-size dependence of errors of the Polyakov-loop expectation values for $N=5$ and $\beta=1.6,1.8$.
For $\beta=1.6$, the plateau appears around $N_{\rm bin}=160000$.
For $\beta=1.8$, the plateau does not appear within the number of samples we adopt in this work.
(For $\beta=1.7$, we cannot judge whether or not the plateau exists due to the fluctuation.)
As with the case of $N=3$, we roughly estimate the simulations for $\beta\leq 1.6$ are performed with sufficient number of samples beyond the auto-correlation times while they are insufficient for higher $\beta$.

We have done the same analysis for $N=10,20$.
We roughly estimate that the simulations for $N=10,20$ are performed with sufficient numbers of samples for $\beta\leq 1.5$ as shown in FIg.~\ref{fig:N_statP} with the gray shade.

\section{Thermal entropy density from the free energy density}
\label{sec:free_energy}

In this appendix, we derive thermal entropy density for a single free scalar field with the TBC.
The TBC for a scalar field $\phi$ is given by
\beq
\phi(x,\tau+L_{\tau}) = e^{i\alpha} \phi(x,\tau)\,.
\eeq
Then, the TBC analog of the partition function is given by
\beq
Z\,=\, \prod_{k_{\tau},k_{s} = -\infty}^{\infty} {1\over{ \left( \frac{2\pi k_{\tau} +\alpha}{L_{\tau}} \right)^{2} + \left(\frac{2\pi k_{s}}{L_{s}}\right)^{2} + m^{2} }}\,.
\eeq
By using the relation 
\beq
\sum_{k=\infty}^{\infty} \delta (p-2\pi k/L) = {L\over{2\pi}} \sum_{n=-\infty}^{\infty} e^{i n L p},
\eeq
the TBC analog of the free energy 
$F=-{1\over{L_{\tau}}} \log Z$ can be rewritten as
\beq
F \,=\,  {\sum}' L_{s} \int {d^{2} p \over{(2\pi)^{2}}} e^{i(n_{\tau} (L_{\tau} p_{\tau} -\alpha) + n_{s} L_{s} p_{s})} \log(p^{2} + m^{2}) + const\,,
\eeq
where $\sum'$ denotes the summation over $(n_\tau, n_s) \in \Z^2$ excluding the term with $(n_\tau,n_s) = (0,0)$ 
(denoted as $const$). 
Performing the integration, we obtain 
\beq
F \,=\, - {\sum}' e^{-i n_{\tau} \alpha} \frac{mL_{s}}{\pi\sqrt{L_{\tau}^{2} n_{\tau}^{2} + L_{s}^{2} n_{s}^{2}}} K_{1}(m\sqrt{L_{\tau}^{2} n_{\tau}^{2} + L_{s}^{2} n_{s}^{2}})\,,
\eeq
where $K_{1}$ stands for the modified Bessel function
of the second kind.
In the large $L_{s}$ regime, the leading contribution of the free energy density is given by the terms with $n_{s}=0$,
\beq
f \,=\, \lim_{L_{s}\to \infty} {F\over{L_{s}}} = \sum_{n_{\tau}=1}^{\infty} \frac{4m \cos n_{\tau} \alpha}{\pi L_{\tau} n_{\tau}} K_{1}(mL_{\tau} n_{\tau})\,. 
\eeq
For small $L_{\tau}$, we obtain
\beq
f = -  \sum_{n_{\tau}=1}^{\infty} \frac{2 \cos n_{\tau} \alpha}{\pi L_{\tau}^{2} n_{\tau}^{2}}\,+\, {\mathcal O}(m/L_{\tau})\,,
\eeq
and
\beq
\epsilon = {\partial \over{\partial L_{\tau}}}(L_{\tau} f) = \sum_{n_{\tau} =1}^{\infty}  \frac{2 \cos n_{\tau} \alpha}{\pi L_{\tau}^{2} n_{\tau}^{2}}\,+\, {\mathcal O}(m/L_{\tau})\,.
\eeq
From these results, we obtain
\beq
s = L_{\tau}(\epsilon - f) =  \sum_{n_{\tau} =1}^{\infty}  \frac{4 \cos n_{\tau} \alpha}{\pi L_{\tau} n_{\tau}^{2}}\,+\, {\mathcal O}(m)\,.
\eeq
By taking the twist phase as $\alpha=2\pi a/N$ and summing over 
$a=1, \cdots, N$, we obtain the pseudo-entropy for a single scalar field 
\beq
s = -{2\pi\over{3L_{\tau} N}} + \mathcal O(m).
\eeq


\end{document}